\documentclass[fleqn,usenatbib]{mnras}
\usepackage{amssymb,amsmath,graphicx,hyperref}
\usepackage{color}
\usepackage{enumerate}
\usepackage{booktabs}
\usepackage{tabularx}

\newcommand{\hii}         {\mbox{\rm H{\small II}}}
\newcommand{\hi}         {\mbox{\rm H{\small I}}}
\newcommand{\htwo}        {\mbox{H$_{2}$}}

\newcommand{\Sgas}        {\mbox{$\Sigma_{\rm gas}$}\,}
\newcommand{\Smol}        {\mbox{$\Sigma_{\rm mol}$}\,}
\newcommand{\Shi}        {\mbox{$\Sigma_{\rm H {\small I}}$}\,}

\newcommand{\Sstar}       {\mbox{$\Sigma_*$}}
\newcommand{\Ssfr}        {\mbox{$\Sigma_{\rm SFR}$}\,}

\newcommand{\Phyd}        {\mbox{P$_{\rm h}$}}
\newcommand{\Pde}        {\mbox{P$_{\rm DE}$}}

\newcommand{\ha}          {\mbox{H$\alpha$}}
\newcommand{\hb}          {\mbox{H$\beta$}}
\newcommand{\msunperpcsq} {\mbox{\rm M$_\odot$~pc$^{-2}$}}

\newcommand{\msunperyrkpcsq} {\mbox{\rm M$_\odot$~yr$^{-1}$ kpc$^{-2}$}}

\newcommand{\DrSFMS}{\mbox{$\Delta {\rm SFMS}$}}
\newcommand{\DrSFE}{\mbox{$\Delta {\rm SK}$}}
\newcommand{\Drfmol}{\mbox{$\Delta {\rm MGMS}$}}
\newcommand{\DSFHP}{\mbox{$\Delta {\rm SFHP}$}}

\title[Star-formation regulation at kpc scales]{EDGE-CALIFA survey: Self-regulation of Star formation at kpc scales.}

\author[J.K. Barrera-Ballesteros  et al.]{J.K. Barrera-Ballesteros$^{1}$,\thanks{e-mail: jkbarrerab@astro.unam.mx} 
S.F. S\'anchez$^{1}$, 
T. Heckman$^{2}$, 
T. Wong$^{3}$, 
A. Bolatto$^{4}$, 
E. Ostriker$^{5}$, \newauthor 
E. Rosolowsky$^{6}$,
L. Carigi$^{1}$, 
S. Vogel$^{4}$,
R. C. Levy$^{4}$, 
D. Colombo$^{7}$, 
Yufeng Luo$^{3}$, 
Yixian Cao$^{3,8}$ \newauthor 
\& the EDGE-CALIFA team
\\
$^{1}$Instituto de Astronom\'ia, Universidad Nacional Aut\'onoma de M\'exico, A.P. 70-264, 04510 M\'exico, D.F., M\'exico \\
$^{2}$Department of Physics \& Astronomy, Johns Hopkins University, Bloomberg Center, 3400 N. Charles St., Baltimore, MD 21218, USA \\
$^{3}$Department of Astronomy, University of Illinois, Urbana, IL 61801, USA \\
$^{4}$Department of Astronomy, University of Maryland, College Park, MD 20742, USA \\
$^{5}$Department of Astrophysical Sciences, University of Princeton, Peyton Hall, 4 Ivy Lane, Princeton, NJ 08544, USA \\
$^{6}$Department of Physics, University of Alberta, 4-181 CCIS, Edmonton, AB T6G 2E1, Canada \\
$^{7}$Max-Planck-Institut f\"ur Radioastronomie, Auf dem H\"ugel 69, 53121 Bonn, Germany \\
$^{8}$Aix Marseille Université, CNRS, LAM (Laboratoire d’Astrophysique de Marseille), F-13388 Marseille, France
}

\date{Accepted XXX. Received YYY; in original form ZZZ}

\pubyear{2020}

\begin{document}
\label{firstpage}
\pagerange{\pageref{firstpage}--\pageref{lastpage}}
\maketitle

\begin{abstract}
The processes that regulate star formation in the local Universe at kpc scales are essential to understand how galaxies evolve. We present the relation between the star formation rate density, \Ssfr, and the hydrostatic midplane pressure, \Phyd\ , for 4260 star-forming regions of kpc size located in 96 galaxies included in the EDGE-CALIFA survey covering a wide range of stellar masses and morphologies. We find that these two parameters are tightly correlated, exhibiting smaller scatter and strong correlation in comparison to other star-forming scaling relations. A power-law, with a slightly sub-linear index, is a good representation of this relation. Locally, the residuals of this correlation show a significant anti-correlation with both the stellar age and metallicity whereas the total stellar mass may also play a secondary role in shaping the \mbox{\Ssfr - \Phyd} relation. For our sample of active star-forming regions (i.e., regions with large values of \ha\, equivalent width), we find that the effective feedback momentum per unit stellar mass ($p_{\ast}/m_{\ast}$), measured from the \Phyd/\Ssfr\, ratio increases with \Phyd. The median value of this ratio for all the sampled regions is larger than the expected momentum just from supernovae explosions. Morphology of the galaxies, including bars, does not seem to have a significant impact in the \mbox{\Ssfr - \Phyd} relation. Our analysis suggests that self regulation of the \Ssfr at kpc scales comes mainly from momentum injection to the interstellar medium from supernovae explosions. However, other mechanism in disk galaxies may also play a significant role in shaping the \Ssfr at local scales. Our results also suggest that \Phyd\, can be considered as the main parameter that modulates star formation at kpc scales, rather than individual components of the baryonic mass.
\end{abstract}




\section{Introduction}
\label{sec:Intro}

The understanding of the physical conditions that lead to the formation of new stars in galaxies is essential to understand their formation and evolution. There are basically two physical scenarios that drive the star formation in disk galaxies \citep[see a review in][]{Kennicutt_2012}. In the first scenario, the star formation is mostly controlled by the properties and amount of the interstellar medium (ISM). In this so-called 'bottom-up' picture, the local star formation rate is controlled completely by the amount of dense gas and structure of the molecular clouds \citep[e.g.,][]{Krumholz_2005}. On the other hand, in the so-called 'top-down' scenario, local star-formation is controlled largely by global dynamical events and dynamical timescales \citep[e.g.,][]{Silk_1997}. In the latter scenario, variations of star-formation is controlled by gravitational instabilities in the disk rather than cooling of molecular clouds, with no distinction between densities regimes of the molecular clouds that can affect the amount of newly formed stars. Another model that emerge from these two views is the self-regulation star-formation scenario in which the hydrostatic pressure from the baryonic component balances the feedback from newly formed massive stars reaching an equilibrium \citep[e.g., ][]{Cox_1981,Dopita_1985,Silk_1997}. 

Using data from the PHANGS survey, \cite{Sun_2020} found that the dynamical pressure (i.e., the pressure due to self-gravity and external disk gravity) is in equilibrium with the turbulent pressure for most of their sampled molecular clouds located in nearby star-forming galaxies. Different mechanisms can be considered that could explain such equilibrium including momentum flux injection to the interstellar medium (ISM) from stellar feedback \citep[including supernovae explosions, stellar winds and radiation,  e.g.,][]{Thompson_2005, Ostriker_Shetty_2011,Faucher-Giguere_2013} or/and gravitational instabilities \citep[e.g., ][]{Ibanez_2017,Krumholz_2016}. The injection of momentum due to stellar feedback into the ISM is a rather sporadic and very localized event that may not occur in every single location of the galaxy where molecular gas is available. Therefore, in the self-regulated framework, a star-forming galaxy can be considered as a quasi-steady-state system \citep{Ostriker_2010,Ostriker_Shetty_2011}. The equilibrium between the pressure and the star-formation feedback has to be considered on spatial scales significantly larger than the typical size of giant molecular clouds (few tenths of pc) and temporal scales larger than a cycle of star formation. Numerical simulations suggest that these scales are of the order of few kpc and few hundreds of Myr \citep[e.g.,][]{Kim_2017,Semenov_2017, Orr_2018}.

Recently, the relation between the star formation surface density and the pressure at kpc scales has been investigated extensively in different surveys including star-forming galaxies. Using a sample of 23 dwarf and disk galaxies included in the HERACLES survey, \cite{Leroy_2008} found a strong correlation between these two observables for radial bins. \cite{Herrera-Camus_2017} and \cite{Sun_2020} found similar results using spatially resolved dataset from samples of 31 and 28 star-forming galaxies in the nearby Universe ($D \lesssim 30$ Mpc) included in the KINGFISH and PHANGS surveys, respectively. In general, the strong correlation between the star formation surface density (\Ssfr) and the pressure is in agreement with a linear relation (i.e., power-law index close to one). Theory  and numerical simulations suggest that this could be the case in the scenario in which supernovae feedback is the main supplier of pressure against the hydrostatic pressure \citep[e.g.,][]{Ostriker_Shetty_2011, Kim_2013}.   

At those kpc scales \Ssfr exhibits a strong correlation with the surface gas density (\Sgas) in star-forming galaxies \citep[also known as Schmidt-Kennicutt relation, see, rSK,][ and reference therein]{Kennicutt_2012}. On the other hand, Integral Field Spectroscopy (IFS) used in large samples of star-forming galaxies has make clear the strong relation between \Ssfr and the stellar component of the baryonic mass at kpc scales, \Sstar\, known as the resolved star-formation main sequence \citep[rSFMS, see a review in ][and references therein]{Sanchez_2020ARAA}. Both baryonic components, tracing the local gravitational potential or similarly the hydrostatic pressure, may provide a better estimate of \Ssfr at kpc scales (e.g., Barrera-Ballesteros et al., submitted). Therefore we would like to explore whether a star-forming scaling relation that uses both components of the baryonic mass (like the hydrostatic mid-plane pressure, \Phyd) provides a better description of \Ssfr than those relations using individual components of the baryonic mass.

Among the recent efforts to explore the local properties of galaxies in the nearby Universe, the CARMA Extragalactic Database for Galaxy Evolution (EDGE) survey \citep{Bolatto_2017} has mapped the molecular gas in 126 galaxies observed IFS data from the Calar Alto Legacy Integral Field Area (CALIFA) survey \citep{Sanchez_2012}. This yields a dataset that allow us not only to study the interplay of the molecular gas with other observables derived from the optical spectra at kpc scales, but also to explore the impact of the global/integrated properties in the derived local scaling relations. The EDGE-CALIFA survey overcomes the so-called "cosmic variance" problems, meaning that it samples a sufficient volume of the Universe to truly represent the relations at local scales \citep[e.g.,][]{Diemer_2019}. Among other results, this survey has improved our understanding of how the molecular gas depletion time changes across the extension of galaxies \citep{Utomo_2017}, on how different local and global parameters affect \Ssfr using a multi-linear approach \citep{Dey_2019}, and on how we can characterize the molecular gas at kpc scales using the optical extinction \citep{BB2020}. Using the EDGE-CALIFA spatially resolved dataset, we explore in this article the relation between \Ssfr and \Phyd. We find that these two parameters are strongly correlated, suggesting that star-formation at kpc scales, in a significant fraction of regions located in galaxies in the nearby Universe, is consistent with the self-regulation scenario \citep[e.g., ]{Elmegreen_1989,Silk_1997}. We explore the impact of local and global observables in this relation as well as its role in comparison to other star-forming scaling relations. This paper is organized as follows. In Sec.~\ref{sec:Data} we present the spatially-resolved and ancillary data used in this study which is available in the \texttt{edge\_pydb} database (Wong et al. in prep.). In Sec.~\ref{sec:Results} we show the main results of this paper. In Sec.~\ref{sec:Disc} we discuss these results. Finally, in Sec.~\ref{sec:Conclusions} we present the main conclusions of this article. 

\section{Data and Analysis}
\label{sec:Data}
\begin{figure*}
\includegraphics[width=\linewidth]{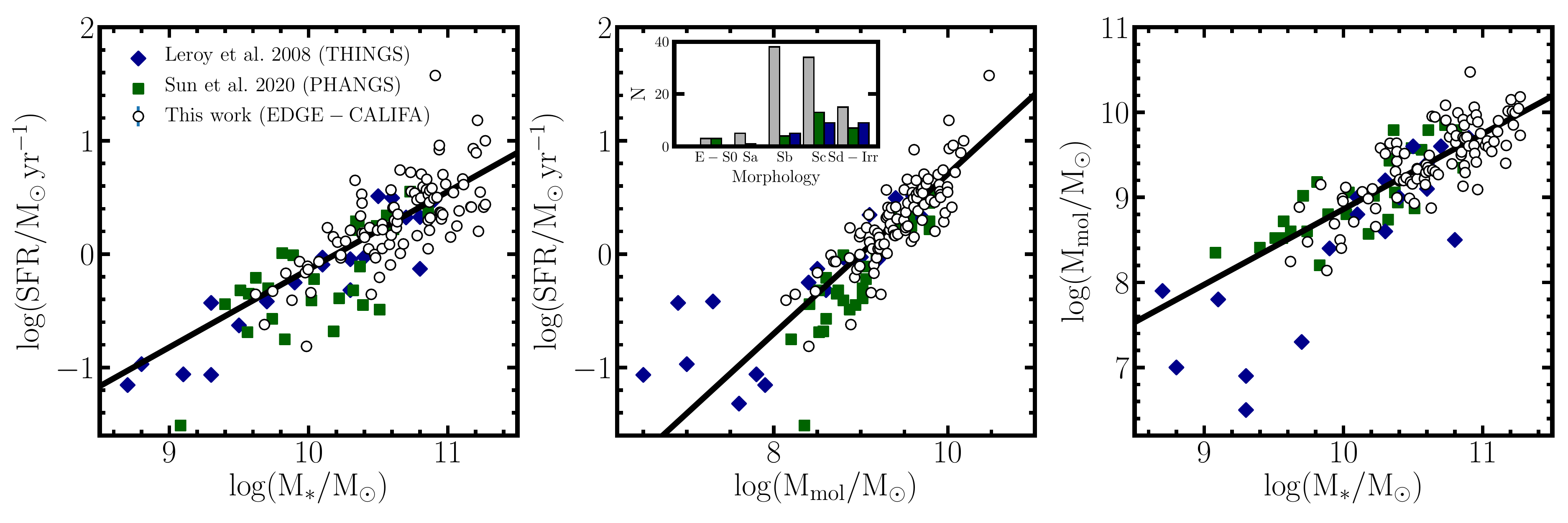}
\caption{Comparison of the 96 EDGE-CALIFA galaxies used in this study (empty circles) with different samples where the spatially-resolved \mbox{\Ssfr - \Phyd} relation has been derived. Blue diamonds, and green squares represent the samples from the THINGS \citep{Leroy_2008}, and PHANGS surveys \citep{Sun_2020}, respectively. The samples are presented in the SFR-M$_\mathrm{\ast}$ (left panel), the SFR-M$_\mathrm{mol}$ (middle panel), and the M$_\mathrm{\ast}$-M$_\mathrm{mol}$ planes (right panel). The solid lines represent the best linear fits to the data points in our sample. The inset in the middle panels shows the distribution of morphologies for this sample (empty bins), the THINGS+HERACLES (blue bins) and the PHANGS (green bins) surveys, respectively. Our sample covers a significant dynamic range in global properties for a variety of morphological types, which is essential to study the impact of global parameters on the \mbox{\Ssfr - \Phyd} relation at kpc scales.
}
\label{fig:Sample}
\end{figure*}
\subsection{The CALIFA and EDGE surveys}
\label{sec:CALIFA-EDGE}
Here we provide a brief description of the optical Integral Field Spectroscopy (IFS) CALIFA and the molecular gas EDGE surveys, respectively. Observations from both surveys comprise the core datasets that we use in this study to perform our analysis.   

The CALIFA (Calar Alto Legacy Integral Field Area) survey \citep{Sanchez_2012} provides IFS data for more than 600 galaxies in the nearby Universe (0.005$< z < $0.03) using the PMAS/PPAK Integral Field Unit (IFU) instrument \citep{Roth_2005} mounted at the 3.5 m telescope of the Calar Alto Observatory. The instrument consists of 331 fibers of 2\farcs7 diameter each, concentrated in a single hexagon bundle covering a field-of-view (FoV) of $\sim$ 1 arcsec$^2$ with a filling factor of $\sim$ 60\%\,. To provide a full coverage of the FoV a three-point dithering is performed. The average resolution of this instrument is $\lambda$/$\Delta\lambda$\,$\sim$\,850 at $\sim$5000\AA\, with a typical wavelength range from  3745 to 7300\AA. The isophotal diameter of CALIFA galaxies are in the range 45 $\lesssim D_{25}\lesssim$\,80 arcsec in the SDSS (Sloan Digital Sky Survey) $r$-band \citep{Walcher_2014}. The data reduction is performed by a pipeline designed specifically for the CALIFA survey. The final data cube for each galaxy consists of more than 5000 spectra with a sampling of 1$\times$1 arcsec$^2$ per spaxel. The reduction process is described in detail in \cite{Sanchez_2012}, improvements on the reduction  pipeline as well as extensions to the original sample (reaching a total of 834 galaxies) are presented by \cite{Husemann_2013,GarciaBenito_2015,Sanchez_2016}.

The EDGE survey obtained millimeter-wave interferometric observations for a sample of 126 galaxies included in the CALIFA survey. These observations were carried out at the Combined Array for Millimeterwave Astronomy \citep[CARMA, ][]{Bock_2006}. The EDGE survey provides the first effort to combine resolved CO data with IFS optical data for a significant sample of galaxies representative of the local Universe. We present a brief description of the survey here, see \cite{Bolatto_2017} for a detailed description. Galaxies were observed using half-beam-spaced seven-point hexagonal mosaics yielding a half-power field-of-view of radius $\sim$ 50\arcsec. Each galaxy has been observed in both the E and D-array configuration with integration times typically of 40 min and 3.5 hr, respectively. The typical resolution for each configuration is 8 and 4 arcsecs, respectively. The final maps combined the E and D array observations resulting in a velocity resolution of 20 km s$^{-1}$ with a typical angular resolution of 4.5\arcsec and typical rms sensitivity of 30 mK at the velocity resolution. Assuming a Milky-Way constant CO-to-\htwo\, conversion factor \citep[$\alpha_{\rm CO} = 4.3\,{\rm M_{\odot} (K\,km\,s^{-1}\,pc^{2})^{-1}} $,][]{Bolatto_2013}, the survey is sensitive to an \htwo\, surface mass density of  $\sim\,4-110\,\,\mathrm{M_{\odot}\,\,pc^{-2}}$ (averaged over a $\sim$ 1.5 kpc scale). The data cubes are smoothed and then masked in order to distinguish CO signal from noise and to reach higher signal to noise \citep[see more details in][]{Bolatto_2017}.

\subsection{The \texttt{edge\_pydb} database}
\label{sec:edgepy}

As we show above, the spatially resolved dataset from the IFU CALIFA survey and the dataset from the CO millimeter obtained from CARMA array are relatively different in terms of field-of-view coverage, spatial resolutions and spatial sampling. The \texttt{edge\_pydb} database has been created as a homogeneous source of optical and millimeter maps and data for the 126 EDGE galaxies to be used in a flexible python environment that allows a simple yet robust exploration of the EDGE-CALIFA dataset. This database also provides an integration of external properties with the spatially resolved information. A detailed description of the database can be found in Wong et al. (in prep.). Here we highlight its main features and the data used in this analysis.

This database provides different estimates of the CO moments from the CARMA observations. The database also provides a smoothed and masked version of the CARMA CO datacubes. The CO datacubes are integrated in the velocity axis to obtain the surface brightness maps (\texttt{smo} table). Both CALIFA and EDGE datasets are convolved to the same  spatial resolution (i.e., 7\arcsec). Then the \textsc{Pipe3D} data analysis pipeline \citep{Sanchez_2015, Sanchez_2016_pipe3d} is run over the convolved optical datacubes resulting in two dimensional maps of optical properties with the same resolution as the CO surface brightness maps. By fitting the stellar continuum (using a single stellar population fitting, SSP, adopting a \cite{Salpeter_1955} initial Mass Function, IMF) and the emission lines for each of the spaxels in each datacube, this pipeline extracts two dimensional maps of a given stellar or ionized gas observable. The maps are sampled on a square grid spaced by 3\arcsec\, in RA and DEC after interpolation to the CARMA WCS. The grid is shifted so that the reference pixel is retained. The tables with ancillary data include information from the LEDA and NED databases, IR photometry from the WISE survey among others. The database also include the information from integrated properties derived from the CO and optical datacubes. For each of the galaxies in this study,  we use the values from the database of: the total molecular gas mass, (M$_\mathrm{mol}$), the total stellar mass (M$_\mathrm{\ast}$), the integrated star formation rate (SFR), the effective radius (R$_{\rm eff}$), the stellar scale height ($l_s$), the minor-major axis ratio ($b/a$), and the morphology. The reader is addressed to Wong et al. (in prep.) for a detailed description on how these observables are derived or from which database they have been obtained.   

\subsection{Derived Quantities}
\label{sec:Quants}

We use the maps of the molecular gas density, \Smol, for each galaxy from the \texttt{edge\_pydb} database. These maps are obtained by converting the CO surface brightness maps from the \texttt{smo} list into molecular gas mass density maps using a constant CO-to-\htwo\, conversion factor following \cite[][$\alpha_{\rm CO} = 4.3\,{\rm M_{\odot} (K\,km\,s^{-1}\,pc^{2})^{-1}}$]{Bolatto_2013}. This factor includes the mass contribution from helium (below we also estimate a variable conversion factor). 

The \texttt{edge\_pydb} database provides all the maps from both the stellar continuum and the ionized gas components derived from the \textsc{Pipe3D} pipeline. From the fitting of the stellar continuum we use the stellar surface mass density map (\Sstar), the stellar age and metallicity ([Z/H]), and the stellar velocity dispersion ($\sigma_{\ast}$) for each of the sampled regions. From the analysis of emission lines of the ionized gas we use the integrated flux maps of \ha, \hb, \mbox{\rm [O{\small III}]} and \mbox{\rm [N{\small II}]} emission lines. We also use the equivalent width map of the \ha\, emission line \citep[EW(\ha), see details in ][]{Sanchez_2016_pipe3d}. From these emission line fluxes we derive the Balmer decrement ratio (\ha/\hb). From this ratio and the \ha\, emission line, we obtain the extinction-corrected star formation rate surface density map, \Ssfr \citep{Kennicutt_1998_SFR}. All the surface densities are corrected by the galaxy's inclination following \cite{BB_2016}. For this study, we adapted both the \Ssfr, and \Sstar densities to a Kroupa IMF. This is equivalent to multiply these quantities in the database (Salpeter IMF) by a factor of 0.61 \citep[see][review]{Madau_Dickinson_2014}. As proxy for the inclination we use the $b/a$ axis ratio. The typical relative error of \Smol, \Sstar, and \Ssfr\, are $\sim$ 0.28, 0.15, and 0.20 dex, respectively. 

We follow \cite{Elmegreen_1989} to derive for each region the mid-plane hydrostatic pressure\footnote{Note that we use the traditional term `hydrostatic', even though the gas in the ISM is not static, with the majority of the pressure associated  with turbulence} (\Phyd):
\begin{equation}
\label{eq:Phyd}
    \Phyd = \frac{\pi}{2}\, G\, \Sgas \left( \Sgas + \frac{\sigma_{\rm gas}}{\sigma_{\rm \ast,z}}\, \Sstar \right),
\end{equation}
where $G$ is the gravitational constant, \Sgas and \mbox{$\sigma_{\rm gas}$} are the total gas surface mass density (molecular and atomic \Sgas = \Smol+\Shi) and the total gas velocity dispersion, respectively; and \mbox{$\sigma_{\rm \ast,z}$} is the stellar velocity dispersion in the axis perpendicular to the disk. Since we do not have direct observations of the atomic gas density distribution for these galaxies, to derive the best fit of the \Ssfr - \Phyd\, relation in Sec.~\ref{sec:SPrel} we run a Monte Carlo simulation assuming 1000 realizations with different values of \Shi within a normal distribution centered at \Shi = 7 \msunperpcsq and with a standard deviation of 2  \msunperpcsq. These range of values are representative of the atomic gas densities found in normal star-forming galaxies \citep[e.g.,][]{Bigiel_2008}. The value of \Phyd\, for each region presented in this study is thus the average of the above realizations. We adopt a constant value of the total gas velocity dispersion  \mbox{$\sigma_{\rm gas}$ = 11  km s$^{-1}$}. This value is within the range of typical velocity dispersions found in disk galaxies for both components of the cold gas in the nearby Universe \citep[e.g.,][]{Caldu-Primo_2013, Levy_2018} and have been adopted in different studies of the \mbox{\Ssfr - \Phyd} relation\citep[e.g., ][]{Leroy_2008,Sun_2020}. To estimate \mbox{$\sigma_{\rm \ast,z}$},  we follow \cite{Leroy_2008, Zheng_2013}. By assuming a relation between this dispersion and the disk scale height ($h_s$) and the stellar mass surface density for an isothermal disk \citep{van_der_Kruit_1988} we have
\begin{equation}
\label{eq:sz}
    \sigma_{\rm \ast,z} = \sqrt{2\,\pi\,G\,\Sstar h_s}, 
\end{equation}
here we assume that the EDGE-CALIFA galaxies follow the ratio \mbox{$l_s/h_s$} = 7.3, where $l_s$ is the disk stellar scale length \citep{Kregel_2002, Sun_2020}. In turn, we relate the stellar scale length $l_s$ of the disk to the effective radius of the galaxies in units of pc by adopting \mbox{$l_s$ = R$_{\rm eff}$/1.68} \cite[i.e., assuming a Sersic profile with $n$ = 1, ][]{Graham_2005}. We adopt this relation in order to provide an estimation of $l_s$ in larger samples of galaxies where only \mbox{R$_{\rm eff}$} has been determined. In Appendix \ref{app:Phyd} we compare the \mbox{$\sigma_{\rm \ast,z}$} ratio assuming different estimations of $l_s$. 

In Sec.~\ref{sec:caveats} we study the impact of a variable \mbox{$\mathrm{\alpha_{CO}}$} conversion factor in the estimation of the hydrostatic mid-plane pressure. We use Eq.~7 from \cite{Colombo_2020} to estimate this variable conversion factor: 
\begin{equation}
    \mathrm{\alpha_{CO}}(Z^{\prime},\Sstar) = 2.9 \exp \left(\frac{0.4}{Z^{\prime}}\right) \left( \frac{\Sstar}{100 \msunperpcsq} \right)^{-\gamma},
\label{eq:aCO}    
\end{equation}
where \mbox{$\mathrm{\alpha_{CO}}(Z^{\prime},\Sstar)$} is in units of \mbox{${\rm M_{\odot} (K\,km\,s^{-1}\,pc^{2})^{-1}}$}, \mbox{$Z^{\prime}$} is the ionized gas metallicity relative to the solar one, and $\gamma$ = 0.5 where \mbox{$\Sstar\, > 100 \msunperpcsq$} and $\gamma$ = 0 otherwise.  This is a variation of the variable conversion factor derived in Eq.~31 by \cite{Bolatto_2013}. Following \cite{Colombo_2020}, we assume that the total density in our regions is dominated by \Sstar ($\Sigma_{\rm total} \sim \Sstar$), also that the Giant Molecular Cloud (GMC) molecular gas surface density in units of 100 \msunperpcsq\, is equal to one. The ionized gas metallicity is obtained by using the strong-lines calibrator derived by \cite{Marino_2013}:
\begin{equation}
    12 + \log({\rm O/H})  = 8.533 - 0.214 \left( \frac{{\rm [O{\small III}]}}{\hb} \times \frac{\ha}{{\rm [N{\small II}}]} \right).  
\end{equation}

To provide a reliable comparison with the literature, in Sec.\ref{sec:comp} instead of \Phyd we use a slightly different estimate of the dynamical equilibrium pressure, \Pde. Following \citet{Kim_2011}, and \citet[][]{Sun_2020} \Pde\, is given by:
\begin{equation}
\label{eq:Pde}
    \Pde = \frac{\pi G}{2} \Sigma^2_{\rm gas} +\Sgas\sqrt{2G\rho_{\ast}}\,\sigma_{\rm gas,z} ,
\end{equation}
where $\rho_{\ast}$ is the mid-plane stellar volume density \citep[][see their Eq.~13]{Sun_2020}. Here, $\sigma_{\rm gas, z}$ is the velocity dispersion of the gas perpendicular to the disk, and we adopt a value 11 km/s as above. \cite{Sun_2020} noted that this estimate can slightly increases the measurement of \Pde\, by a factor at most of $\sim$ 0.2 dex in comparison to those dynamical pressures derived directly from their observations. We perform a similar analysis as in  Sec.~\ref{sec:SPrel}, this is, we assume 1000 different realizations of \Shi. We presented in this article the averaged values of \Pde\, from those realizations.

The \texttt{edge\_pydb} database provides the information derived from the CALIFA and EDGE data for over $\sim$16000 individual regions located in the 126 galaxies included the survey. For the analysis in this article we select those star-forming regions with the most reliable estimations of the considered observables. This is, regions where \Smol, \ha, and \hb\, have values larger than three times their errors as well as non-zero estimates of their \Sstar\, and their \Ssfr and EW(\ha)$>$ 6\AA. This selection yields a total of 4260 regions located in 96 galaxies. In Fig.~\ref{fig:Sample} we compare the EDGE-CALIFA galaxies used in this study with other angular-resolved samples used to estimate the \Ssfr - \Pde\, relation in the SFR-M$_\mathrm{\ast}$, in the SFR-M$_\mathrm{mol}$, and in the M$_\mathrm{\ast}$-M$_\mathrm{mol}$ planes. The dynamical range of our sample in these three observables is wide, covering $\sim$ 2 orders of magnitude for each of them. In comparison to our sample, the THINGS sample \citep{Leroy_2008} covers a range of low-mass galaxies whereas the 28 galaxies from the PHANGS sample from \cite{Sun_2020} covers a similar range of properties as our sample \footnote{estimates of M$_\mathrm{mol}$ for the PHANGS survey have been kindly provided by A. K. Leroy (private communication). Full details will be available in Leroy et al. (in prep.). SFR, and M$_\mathrm{\ast}$ values are taken from the website: \texttt{https://sites.google.com/view/phangs/home/sample}}, however the EDGE-CALIFA galaxies includes a wide range in morphologies (see inset in the middle panel of Fig.~\ref{fig:Sample}). In Sec.~\ref{sec:global} we explore how the radial distribution of the \Ssfr - \Phyd\, relation varies for galaxies with different global star-formation activity and global gas fraction derived from the best fits of the above relations (black lines in each panel of Fig.~\ref{fig:Sample}). For our sample of galaxies in the SFR-M$_\mathrm{\ast}$, the SFR-M$_\mathrm{mol}$, and in the M$_\mathrm{\ast}$-M$_\mathrm{mol}$ planes, these fits correspond to: \mbox{$\mathrm{\log(SFR/M_{\odot}\,yr^{-1}) = -7.0 + 0.68\, \log(M_{\ast}/M_{\odot})}$}, \mbox{$\mathrm{\log(SFR/M_{\odot}\,yr^{-1}) = -6.3 + 0.70\, \log(M_{mol}/M_{\odot})}$}, and \mbox{$\mathrm{\log(M_{mol}/M_{\odot}) = 0.88\,\log(M_{\ast}/M_{\odot})}$}, respectively. Furthermore, in Sec.~\ref{sec:comp} we compare the estimate of the \Ssfr - \Pde\, relation using our sample and those derived using samples with resolved measurements. 

\section{Results}
\label{sec:Results}

\subsection{The \Ssfr - \Phyd\, relation at kpc scales for the EDGE-CALIFA galaxies.}
\label{sec:SPrel}
\begin{figure}
\includegraphics[width=\linewidth]{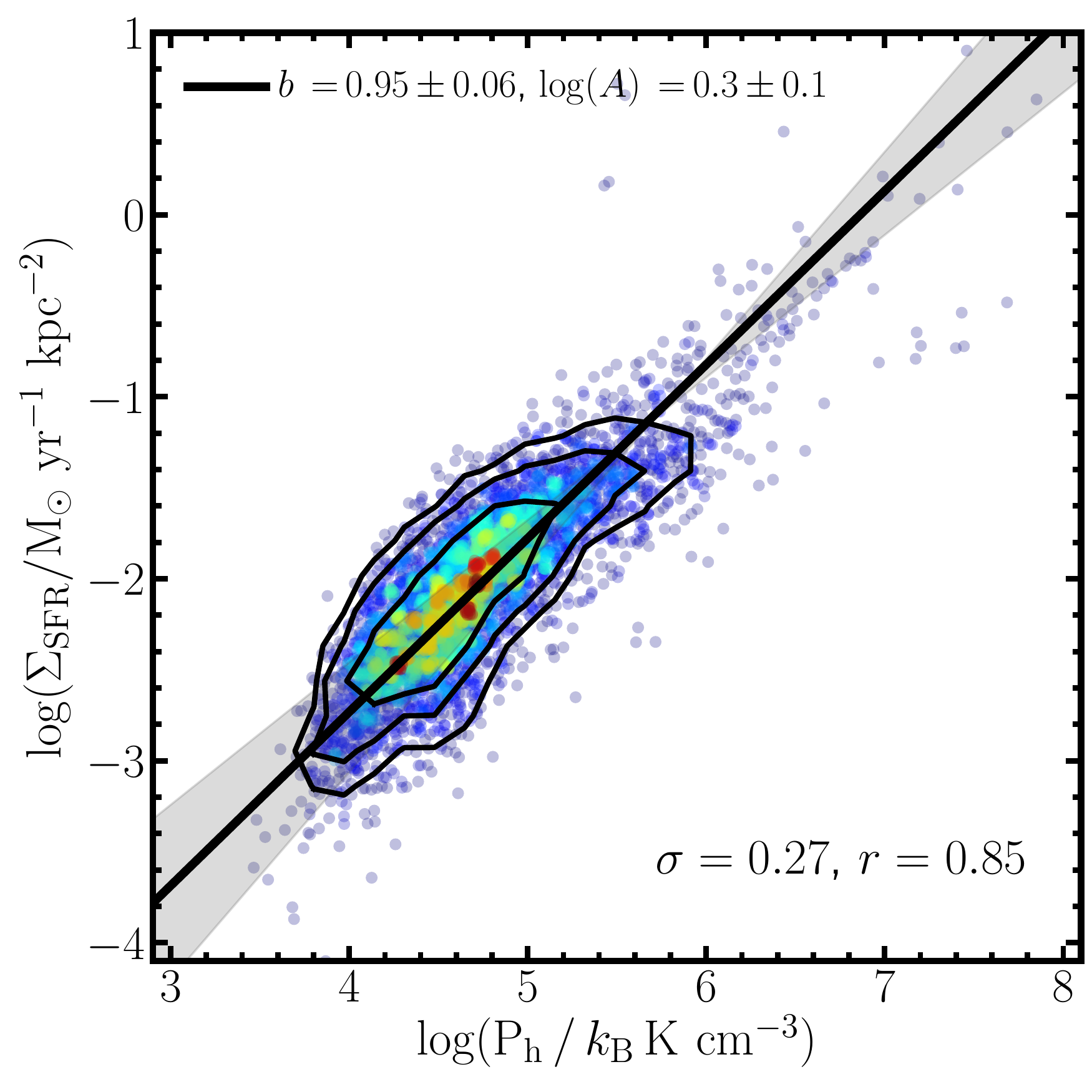}
\caption{
The \Ssfr - \Phyd\, relation for 4260 star forming regions included in 96 galaxies from the EDGE-CALIFA survey. The data points are color-coded according to the density of points. Inwards, the black contours enclose 90\%, 80\% and 50\% of the sample. The solid line represents the best ODR linear fit. The shaded area represents the uncertainty of the fit due to the assumed values of \Shi. The tightness of the relation is indicated by the small standard deviation ($\sigma$), whereas $r$ represents the Pearson correlation coefficient.
}
\label{fig:SPrel}    
\end{figure}

In Fig.~\ref{fig:SPrel} we show the \Ssfr - \Phyd relation for our dataset that includes 4260 regions located in 96 EDGE-CALIFA galaxies. The hydrostatic mid-plane pressure covers $\sim$ four orders of magnitude whereas \Ssfr covers $\sim$ 3 orders of magnitude. On a log-log scale, \Ssfr increases linearly with \Phyd, resulting in a strong correlation (Pearson correlation coefficient of \mbox{$r$ = 0.85}). Similar but smaller correlation coefficients have also been observed in other star-forming scaling relation at kpc scales \citep[e.g., the rSFMS or the rSK, ][]{Cano-Diaz_2016, Lin_2019,Cano-Diaz_2019}. 

To obtain the best parameters that represent the  \Ssfr - \Phyd\, relation we fit the following relation to our dataset using an orthogonal distance regression (ODR) fitting technique: 
\begin{equation}
\label{eq:bestfit}
    \frac{\Ssfr}{10^{-3} \msunperyrkpcsq} =  A \left( \frac{\Phyd}{10^{4}\,\, k_{\rm B}\,\, {\rm K\,\, cm^{-3}}}  \right)^{b},
\end{equation} 
the fitting procedure was repeated 1000 times for each of the different realizations performed to take into account our lack of knowledge of the \Shi distribution (see Sec.~\ref{sec:Quants}). The shaded area in Fig.~\ref{fig:SPrel} represents the best fits from these realizations and the black solid line represents their median. Therefore the reported values of the fit are obtained from this average while their uncertainties come from the scatter of these realizations (\mbox{$b$ = 0.95$\pm$0.05} and \mbox{$A$ = 0.3$\pm$0.1}). We also perform an ordinary least-square (OLS) fit to the average values. We find a slightly flatter relation in comparison to the one derived using the ODR fit (\mbox{$b$ = 0.83$\pm$0.04} and \mbox{$A$ = 0.37$\pm$0.05}). In Sec.~\ref{sec:comp}, we compare these slopes with recent results for regions at sub-kpc scales. The scatter of the residuals  -- measured from their standard deviation, $\sigma$ -- is small compared to other star-forming scaling relations ($\sigma$ = 0.27, see a comparison in Sec.~\ref{sec:local}). We are interested in exploring how other local and global parameters affect this relation. In Secs.~\ref{sec:local} and \ref{sec:global} we study how the residuals from the best fit of this relation correlate with other observables. In Sec.~\ref{sec:Disc} we discuss the possible different scenarios that can explain the sub-linear slope exhibited by this relation.

\begin{figure*}
\includegraphics[width=\linewidth]{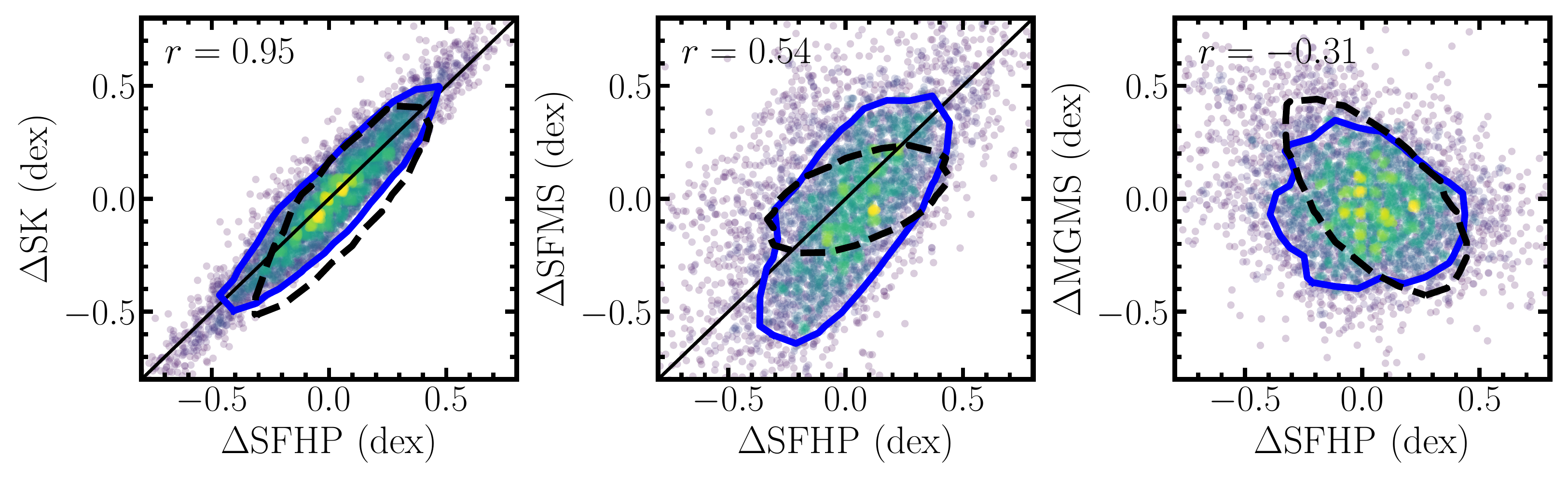}
\caption{The correlation between the residuals of different scaling relations with those derived from the \Ssfr - \Phyd\, relation, \DSFHP. ({\it left})  The residuals of the Schmidt-Kennicutt law (\DrSFE). ({\it middle}) The residuals of the resolved star formation main sequence (\DrSFMS). ({\it right}) The residuals of the molecular gas main sequence (\Drfmol). The number in each panel represents the Pearson correlation coefficient. Solid lines in left and middle panels represent a one-to-one relation. In each panel the blue solid contour encloses $\sim$ 64\% of the sample while the black dashed contour represents the same fraction for those residuals estimated from mock data assuming a constant scatter around best-fit scaling relations (see details in Sec.~\ref{sec:local}). This comparison suggests that correlations between residuals are mostly explained by the strong covariance of the quantities defined in each axis.
}
\label{fig:SPrel_local}    
\end{figure*}
 
\subsection{Impact of Local parameters in the  \Ssfr - \Phyd\, relation}
\label{sec:local}

In the last decade it has become evident that most of the star-forming scaling relations derived for integrated properties are also observable at kpc scales \citep[for a recent review see][]{Sanchez_2020ARAA}. Recent studies have also shown that scatter of the star-forming scaling relations is modulated by different local observables \citep{Ellison_2018,Ellison_2020}. In the previous section we showed that \Ssfr strongly correlates with \Phyd. Furthermore this relation is tight, exhibiting a similar scatter in comparison to other star-forming scaling relations ($\sim$ 0.25 dex). In this section, we explore how the \Ssfr - \Phyd\, relation compares with the three scaling local relations that correlate \Ssfr, \Sstar, and \Smol among them at kpc scales. Then we explore how the \mbox{\Ssfr - \Phyd} relation is modulated by other stellar properties such as the age, metallicity, and their velocity dispersion.  

Using a sample of galaxies included in the ALMAQuest survey, \cite{Lin_2019} found that within star-forming galaxies \Ssfr, \Sstar, and \Sgas closely correlate with each another. On the one hand, \Ssfr correlates with \Sstar \citep[resolved star-formation main sequence, rSFMS; see also ][]{Sanchez_2012, Wuyts_2011,Cano-Diaz_2016, Cano-Diaz_2019}, and with \Smol \citep[resolved Schmidt-Kennicutt relation, rSK; see also ][]{Bigiel_2008}. On the other hand, \Smol\, also correlates with \Sstar \citep[resolved molecular gas main sequence, rMGMS, see also ][]{BB2020}. We derive these scaling relations using our dataset from the EDGE-CALIFA survey in S\'anchez et al. (in prep.). The following are the best relations presented in S\'anchez et al. (in prep.) for this survey:
the resolved Schmidt-Kennicutt relation (rSK):
\begin{equation}
\label{eq:rSK}
    \log \left(\frac{\Ssfr}{\rm M_\odot~yr^{-1} pc^{-2}} \right) = 0.98\,\, \log \left( \frac{\Smol}{\msunperpcsq} \right) -9.01,
\end{equation}
the resolved star formation main sequence (rSFMS):
\begin{equation}
\label{eq:rSFMS}
    \log \left(\frac{\Ssfr}{\rm M_\odot~yr^{-1} pc^{-2}}\right) = 1.02\,\, \log \left( \frac{\Sstar}{\msunperpcsq} \right) - 10.10,
\end{equation}
and the resolved molecular gas main sequence (rMGMS):
\begin{equation}
\label{eq:rMGMS}
   \log \left( \frac{\Smol}{\msunperpcsq} \right) = 0.93\,\, \log \left( \frac{\Sstar}{\msunperpcsq} \right) -0.91,
\end{equation}
their typical scatter is of the order of $\sigma \sim$ 0.25 dex (see details in S\'anchez et al., in prep.). For each of these relation we derive their residuals (i.e., for each star-forming region, the distance in the $y$-direction between the best fit and the data point). In each of the panels in Fig.~\ref{fig:SPrel_local} we plot these residuals against the one derived from the best fit of the \mbox{\Ssfr - \Phyd} relation (\mbox{$\Delta {\rm SFHP}$}): the residuals of the rSK (\DrSFE, left panel),  the residuals of the rSFMS (\DrSFMS, middle panel), and the residuals of the rMGMS (\Drfmol, left panel). In principle, \DrSFE, and \DrSFMS\, quantify the lack (or excess) of \Ssfr for a given \Sgas, and \Sstar, respectively; whereas the \Drfmol\, measures the lack or excess of \Smol for a given \Sstar.

We find that \DrSFMS, and \DrSFE\, correlate with \DSFHP. This is supported by their Pearson correlation coefficients ($r$ = 0.94, and 0.54, respectively). Even more, these two residuals increases linearly with \DSFHP\, following a one-to-one relation (dashed lines in Fig.~\ref{fig:SPrel_local}). This may suggest, that for a given \Phyd the variations in \Ssfr are tightly correlated to those variations expected from  the rSK and rSFMS. The smaller scatter observed in the left panel of  Fig.~\ref{fig:SPrel_local} compare to the one observed in middle panel may indicate that \DrSFE\, has a larger impact in setting \Ssfr for a given \Phyd. In comparison to the previous correlations the residuals of the molecular gas main sequence, \Drfmol, weakly anti-correlates with \DSFHP ($r$ = -0.31, right panel of Fig.~\ref{fig:SPrel_local}). The lack of a strong correlation between these residuals suggests the small impact that the gas fraction has in modulating the star formation at kpc scales in comparison to the stellar and gas surface densities.

Mathematically speaking, the observables from which the previous scaling relations are made of (i.e., the rSK, the rSFMS, and the rMGMS), are the same observables used to derive the \mbox{\Ssfr - \Phyd} relation. Therefore, the correlations among the residuals we find in Fig.~\ref{fig:SPrel_local} can be induced by the strong covariance among those observables. To test the impact of their covariances in the relations of the residuals, in Appendix~\ref{app:Mock} we build the same relations among residuals using a mock dataset by considering the best fits of the scaling relations (i.e., Eqs., \ref{eq:rSK}, \ref{eq:rSFMS}, and \ref{eq:rMGMS}) and assuming the typical scatter from the observables. In each panel of Fig.~\ref{fig:SPrel_local} the blue contour encloses $\sim$64\% of the observed correlation by the residuals whereas the black dashed contours encloses the same fraction from the relations derived from the mock dataset. The comparison between these two distributions shows that the  correlation we find between \DrSFE\, and  \DSFHP\, is mostly driven by the  strong covariance of the observables (left panel of Fig.~\ref{fig:SPrel_local}). The Pearson correlation coefficient derived from the mock dataset is similar as the one obtained from observations ($r$ = 0.85). On the other hand, the distribution of the observed relation between \DrSFMS\, and \DSFHP\, is wider in comparison to one obtained using the mock dataset (middle panel of Fig.~\ref{fig:SPrel_local}). This suggests that the observed correlation between these two residuals is not entirely driven by the covariance of the observables. Finally, the distribution of the observed relation between \Drfmol\, and \DSFHP\, is slightly tighter than the one obtained from the mock dataset (right panel of Fig.~\ref{fig:SPrel_local}). The Pearson correlation coefficient derived from the mock dataset shows a larger anti-correlation between residuals is larger as the one obtained from observations ($r$ = -0.53). The lower correlation coefficient and the smaller scatter observed between \Drfmol\, and \DSFHP\, -- compare with those derived from the mock data -- indicate that \Phyd is an appropriate parameter to describe \Ssfr than the gas fraction. In Sec.~\ref{sec:Disc} we further discuss the impact of individual components of the baryonic mass in the \mbox{\Ssfr - \Phyd} relation.


\begin{figure*}
\includegraphics[width=\linewidth]{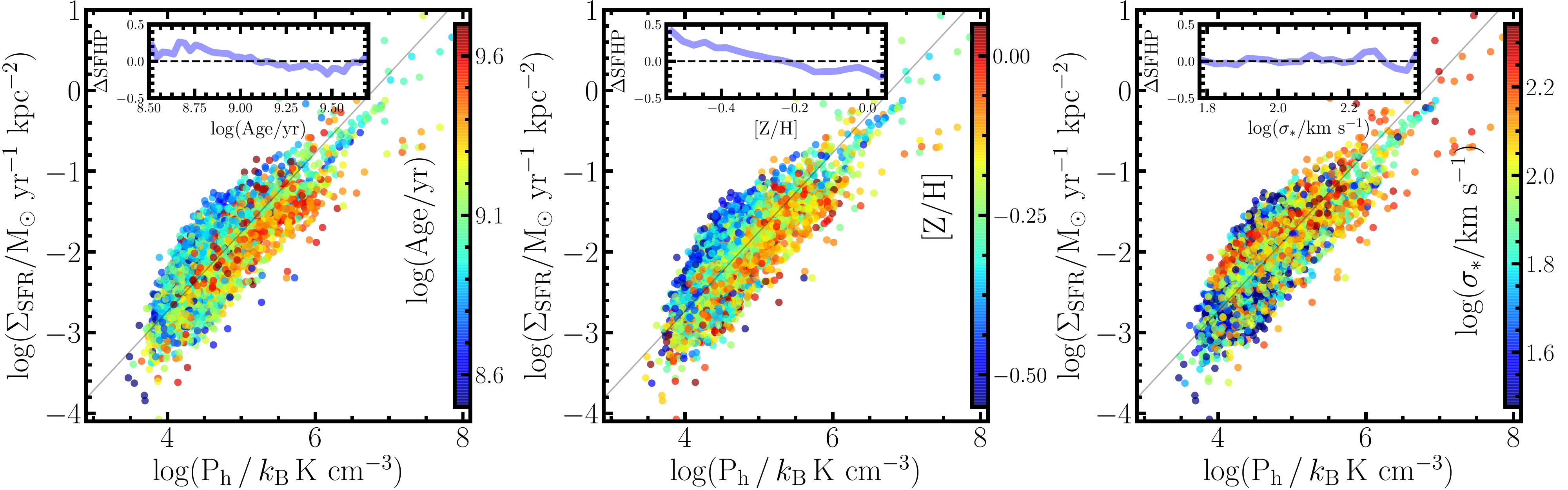}
\caption{The impact of various stellar properties on the \Ssfr - \Phyd\, relation shown in Fig.~\ref{fig:SPrel}.  In the three panels, the insets show the scatter of the relation as a function of the stellar parameter. Left, middle, and right panels show the \Ssfr - \Phyd\, relation color coded by the average stellar age, average stellar age ([Z/H]), and stellar velocity dispersion ($\sigma_{\ast}$), respectively. The residuals of the \Ssfr - \Phyd\, relation (\DSFHP) are anti-correlated with both stellar properties; however \DSFHP\, does not seem to be strongly affected by the stellar velocity dispersion. 
}
\label{fig:SPrel_Z}    
\end{figure*}

Thanks to the \texttt{edge\_pydb} we are able to estimate how stellar properties affect the \mbox{\Ssfr - \Phyd} relation. In the left, middle and right panels of Fig.~\ref{fig:SPrel_Z} we color code Fig.~\ref{fig:SPrel} with respect to the luminosity-weighted stellar age, stellar metallicity ([Z/H]), and the line-of-sight stellar velocity dispersion ($\sigma_{\ast}$), respectively. We find a significant anti-correlation of the stellar age and metallicity with \DSFHP\, ($r$ = -0.25, and  -0.40, respectively). For a given \Phyd, \Ssfr decreases as the age or [Z/H] increases. This suggests that for those regions where the hydrostatic mid-plane pressure is similar, those with young/metal-poor stellar population tend to have large \Ssfr in comparison to those with old/ metal-rich population. These two anti-correlations could emerge from the stellar velocity dispersion. Regions with large/low star formation rate for a given \Phyd\, could be dynamically cold/hot (i.e., low/high $\sigma_{\ast}$). However, in the right panel of  Fig.~\ref{fig:SPrel}, when we color code the \mbox{\Ssfr - \Phyd} relation with respect to $\sigma_{\ast}$, we do not find similar patterns as those observed with the other two stellar properties in the other panels. Even more, the residuals of this relation do not seem to show a correlation with  $\sigma_{\ast}$ ($r$ = 0.01). The spectral resolution of the CALIFA survey only allows to have a reliable estimation of $\sigma_{\ast}$ at high values (i.e.,$\sim$ 60 km s$^{-1}$), but even for large velocity dispersions we do not find a variation of the \DSFHP\, against $\sigma_{\ast}$. 

In Sec.~\ref{sec:global} we find that the scatter of the \mbox{\Ssfr - \Phyd} relation apparently (anti-)correlates with the (total stellar mass) and morphology. In other words, at a fixed \Phyd, \Ssfr\, is higher, for lower ages and metallicities, mainly in low-mass galaxies. Since the stellar metallicity and age are features of the underlying stellar populations, both hint past properties of the ISM, however the \Ssfr\, is a feature of the current ISM. Therefore, it seems that the galactic areas where the younger and metal-poor \hii\, regions are embedded, behave as the outer disk of early-type spiral galaxies (Sa-Sb) or as the inner disk of late-type spiral galaxies (Sc - Sd). These behaviors can be explained by the inside-out formation scenario for the former, and by the outside-in scenario for the latter. In both cases, their star formation histories could be roughly represented as an increasing exponential function. It can also be the case that the observed \DSFHP is affected by our estimation of \Ssfr. The IMF varies with SFR and metallicity, such that very massive stars form when the SFR is high, regardless of the stellar metallicity, but for super solar metallicities the formation of low mass stars dominates, regardless of the SFR. The \Ssfr\, considered in this work is computed using the Kennicutt \Ssfr–\ha\, relation \citep{Kennicutt_1998}. This calibration was obtained by assuming an invariant IMF for stars of solar metallicity with masses between 0.1 and 100 M$_{\odot}$. Therefore this fact may reduce the dispersion of the \mbox{\Ssfr - \Phyd} relation and incidentally it could also modify the slope of other scaling relations such as the rSK and rSFMS \citep[e.g.,][]{Jerabkova_2018}.

We cannot rule out whether these secondary trends could be induce by expected relations between the components of \Phyd and \Ssfr. Although we do not find strong correlations among \Ssfr, \Sstar, \Smol\, and the stellar age and metallicity, we find significant anti-correlations among the specific star formation rate (sSFR = \Ssfr/\Sstar), the star formation efficiency (SFE = \Ssfr/\Smol) and the stellar age and metallicity, respectively. On the other hand, by construction of the \mbox{\Ssfr - \Phyd} relation, \DSFHP\, strongly correlates with both sSFR, and SFE. These inverse relations between sSFR age and metallicity are expected, since low (high) values of sSFR (or equivalently the EW(\ha)) usually trace old/metal-rich (young/metal-poor) stellar populations \citep[e.g.,][]{Mejia-Narvaez_2020}.

\subsection{Impact of Global parameters on the  \Ssfr - \Phyd\, relation}
\label{sec:global}
\begin{figure}
\includegraphics[width=\linewidth]{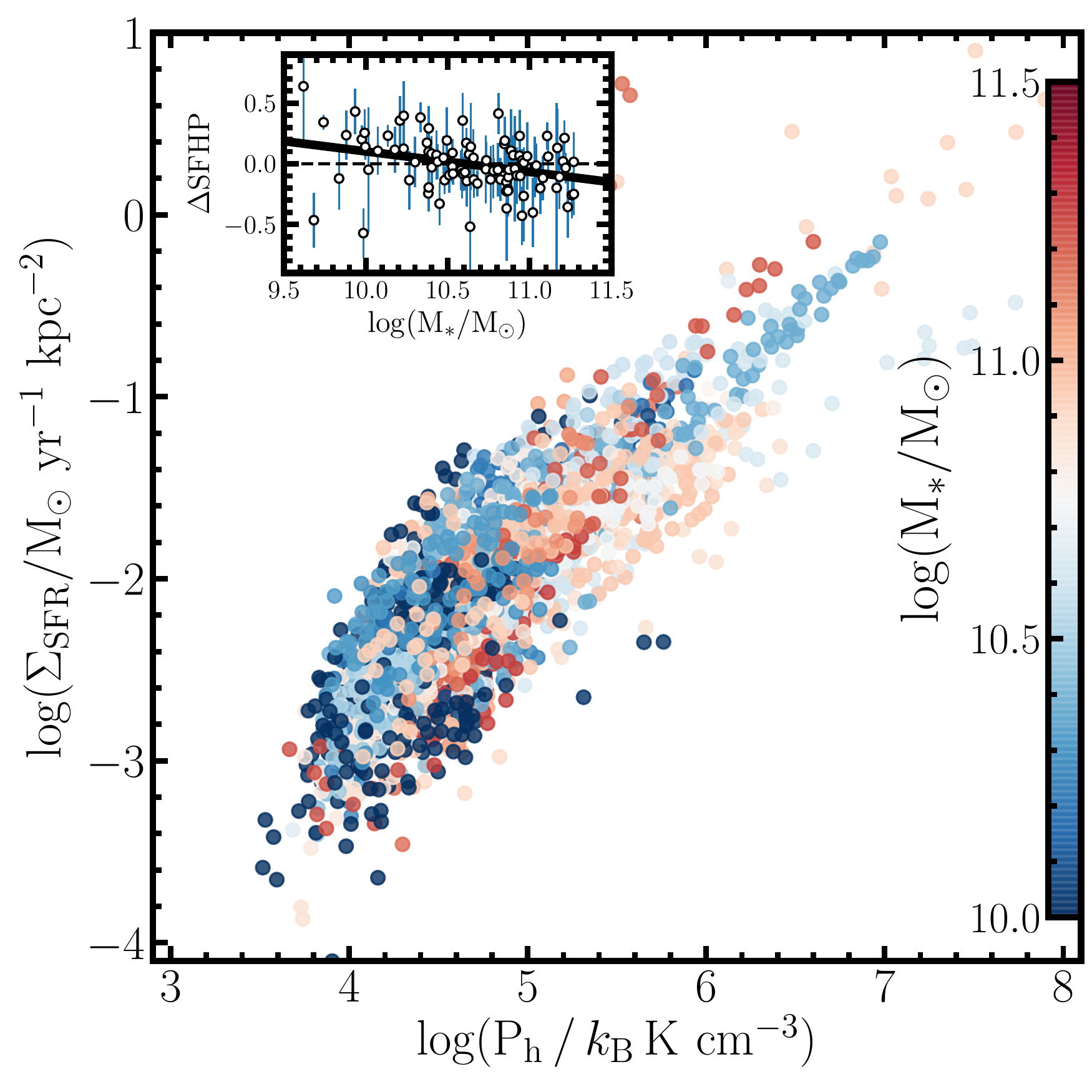}
\includegraphics[width=\linewidth]{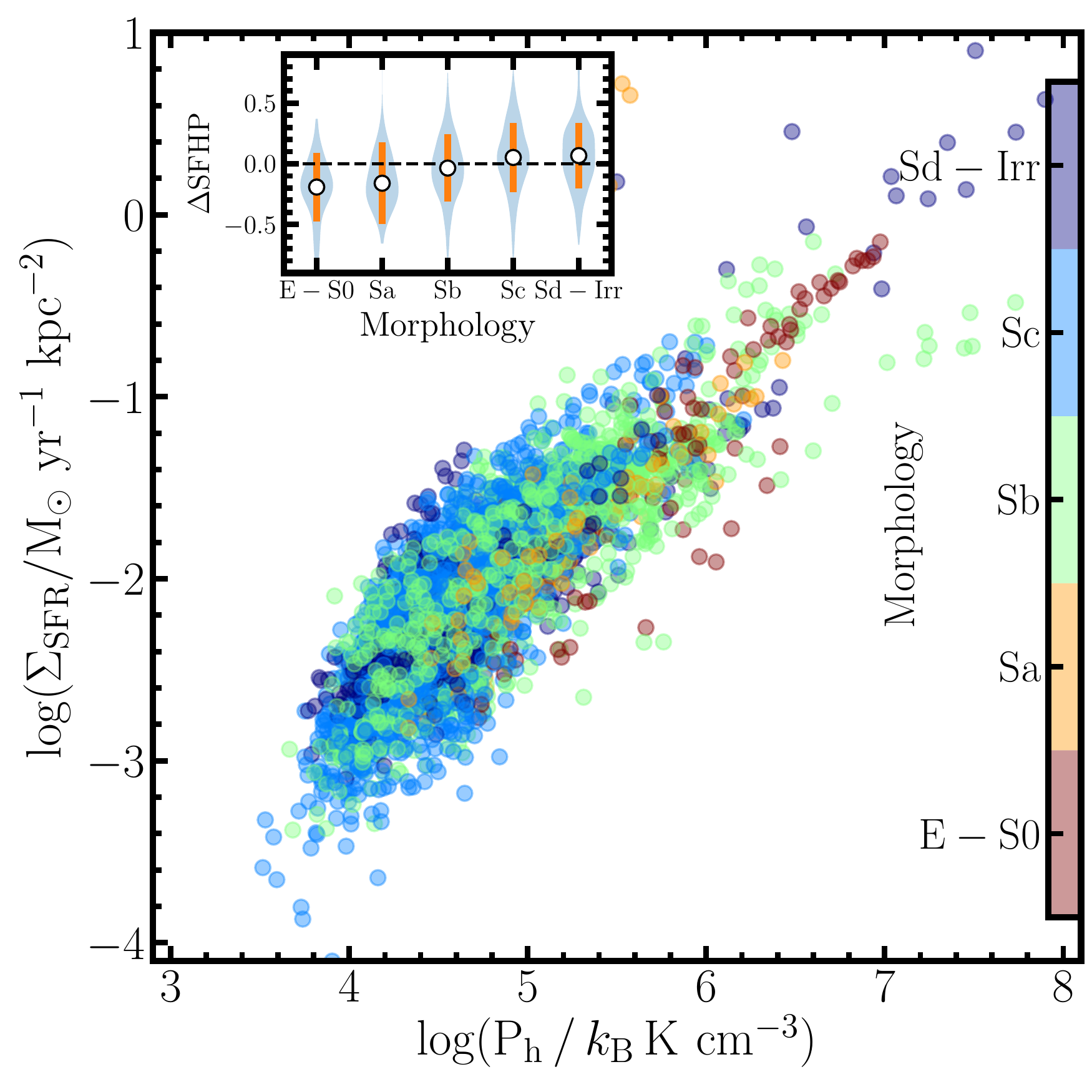}
\caption{({\it top}) The \Ssfr - \Phyd\, relation color coded by the total stellar mass of the host galaxy. The circles in the inset show the average \DSFHP\, for each galaxy against its total stellar mass while the associated error bars represent the standard deviation of \DSFHP\, for each galaxy; the black solid line represents a linear fit to the averaged residuals. ({\it bottom}) The \Ssfr - \Phyd\, relation color coded by the morphological type of the host galaxy. The circles in the inset shows the average \DSFHP\, for a given morphological type. The violin-histograms in the inset represent the distribution of \DSFHP\, for the different morphological types. Although there seems to be a correlation between \DSFHP\, and the total stellar mass, these global properties does not seem to have a large impact in the \Ssfr - \Phyd\, relation (see details in Sec.~\ref{sec:global}.) 
}
\label{fig:Mass_Morph}    
\end{figure}

Having a homogeneous spatially-resolved dataset for a significant large sample of galaxies in the nearby Universe allow us to explore the impact of global properties on the \mbox{\Ssfr - \Phyd} relation. As reviewed by \cite{Sanchez_2020ARAA}, most of the scaling relations derived on kpc scales in the local Universe are affected by the structural/integrated properties of their host galaxy. In particular, the total stellar mass and the morphology of a galaxy can modulate most of these relations at local scales. In Fig.~\ref{fig:Mass_Morph} we color code the \mbox{\Ssfr - \Phyd} relation according to these two global observables. The top panel indicates that for regions located in low-mass galaxies \Ssfr is larger in comparison to regions with similar \Phyd\, located in more massive galaxies. This trend is more evident when we plot the average residuals for each galaxy of the \mbox{\Ssfr - \Phyd} relation against the total stellar mass (see inset). We find that the average \DSFHP\, for each galaxy anti-correlates with the total stellar mass ($r = $ -0.3). For most of low-mass galaxies ($\log({\rm M_{\odot}/M_{\ast}} \lesssim 10.5)$) the best fit of the \mbox{\Ssfr - \Phyd} relation underestimates the observed \Ssfr\, whereas for more massive galaxies \Ssfr\, is slightly overestimated. This is quantified by the slope of the black line in the inset ($\sim -0.17$~dex/M$_{\ast}$). We note that even though this trend seems to be systematic, due to the large scatter, observations for a wider range of galaxies is required to test the robustness of the impact of the total stellar mass in shaping the local \mbox{\Ssfr - \Phyd} relation.

We explore whether the inclusion of the total stellar mass as a secondary parameter in the \mbox{\Ssfr - \Phyd} relation could induce a reduction in its scatter. If so, this would be strong evidence of the importance of the role that the potential well could have in shaping \Ssfr at local scales. We measure the scatter of the relation between \Ssfr and a parameter that includes the stellar mass as a secondary parameter of the form  \mbox{$x = \log(\mathrm{P_{h}}\,/ \,k_\mathrm{B}\,\mathrm{K\,\,cm^{-3} }) - \alpha \,\,\log(\mathrm{M_{\ast}/ 10^{10} M_{\odot}}$)} with  \mbox{$-1 < \alpha < 1$}. We find that the value that yields the smallest scatter is  $\alpha$ = -0.07. The scatter of this relation is very similar to the one where the stellar mass is not included as a secondary parameter (i.e., $\sigma = $ 0.28 dex). We suggest that even though the overall gravitational potential may have an impact in shaping \Ssfr, its impact is relatively mild in comparison to the local pressure. Further studies including larger samples of low-mass galaxies are thus required to quantify the actual impact of the potential well in this relation. 

On the other hand, when we color code this relation according to the morphology of the host galaxies (bottom panel), regions located in late-type galaxies have a slightly large \Ssfr\, than regions in more early-type galaxies for a similar \Phyd. We also identify this mild trend when plotting the average residual of the \mbox{\Ssfr - \Phyd} relation for a given morphological type. For early-type galaxies (E-S0) the best \mbox{\Ssfr - \Phyd} relation overestimates \Ssfr by a factor of $\sim$ 0.2 dex, whereas for late type galaxies (Sc,Sd-Irr) \Ssfr is underestimated by an average of $\sim$ 0.1 dex. Despite this apparent trend we should note that the standard deviation of the residuals for each morphological type (error bars for the white points) are consistent with no change in the residual of the  \mbox{\Ssfr - \Phyd} relation with respect to the morphology (i.e., \DSFHP = 0). Thus to either confirm or rule out these trends we would require a larger sample of galaxies that includes more early-type galaxies than those currently available (see inset in the middle panel of Fig.~\ref{fig:Sample}).   

\begin{figure}
\includegraphics[width=\linewidth]{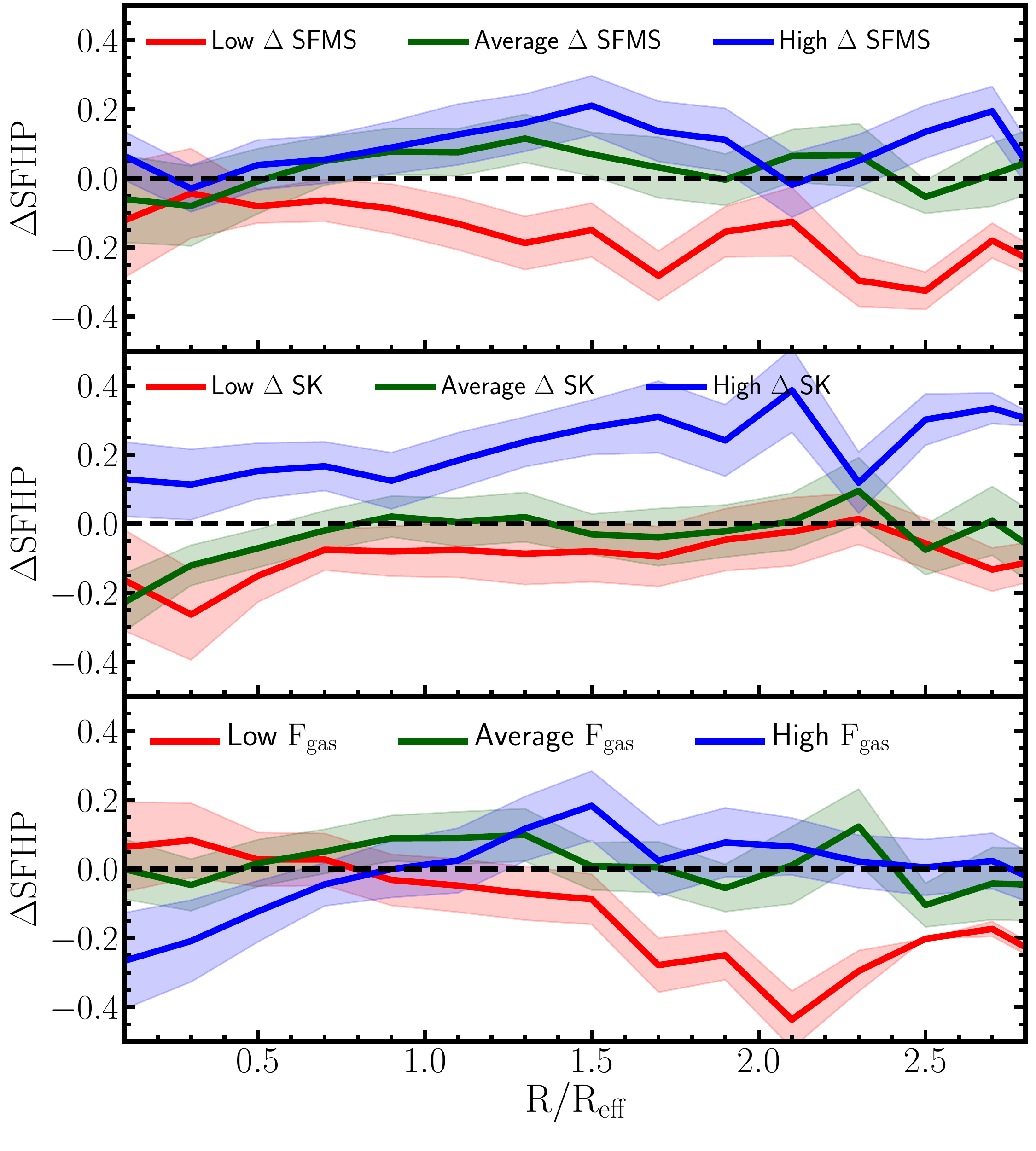}
\caption{The radial distribution of the residuals of the \mbox{\Ssfr - \Phyd} relation for three global scaling relations. For each of the global scaling relation we divide the sample in galaxies with low, average and high bins (red, green, and blue lines) referring to the vertical distance of the galaxy with respect to the best fit of the global scaling relation (see black solid lines in each panel of  Fig.~\ref{fig:Sample}). In the top, middle and bottom panel we show the radial gradients of \DSFHP\, for the star-formation main sequence, the Schmidt-Kennicutt law and the stellar-molecular mass relation, respectively. These radial trends highlight the impact of global properties in the local \mbox{\Ssfr - \Phyd} relation.
}
\label{fig:Rad}    
\end{figure}

In Fig.~\ref{fig:Rad} we plot the radial distribution of the residuals of the \mbox{\Ssfr - \Phyd} relation according to the position of the host galaxy in different global scaling relations presented in Fig.~\ref{fig:Sample}. 
In the top panel of Fig.~\ref{fig:Rad} we plot the radial gradient of \DSFHP\, dividing galaxies in three groups (low, average, and high $\Delta$SFMS) according to their location respect the star formation main sequence (below -0.5, between -0.5 and 0.5, and above 0.5 times the standard deviation of the scatter from the black solid line in the left panel of Fig.~\ref{fig:Sample}, respectively). 
Although the radial distributions are clustered around the zero residual, we find that in galaxies with higher and average SFR for their stellar mass have a flat radial gradient of their \DSFHP.  
For galaxies with low $\Delta$SFMS, \DSFHP\, is negative for all radii, decreasing with radii ($\sim$ -0.07 dex/R$_{\rm eff}$). In other words, for global low star-forming galaxies we are overestimating \Ssfr when deriving it from \Phyd, particularly at they outskirts.

Similar as above, in the middle panel of Fig.~\ref{fig:Rad} we classify galaxies according to their vertical distance with respect to the black solid line in the middle panel of Fig.~\ref{fig:Sample} (low, average, and high $\Delta$SK). Galaxies with low and average SFR for their total gas mass have similar mildly increasing gradients ($\sim$ 0.03 dex/R$_{\rm eff}$). On the other hand, for those galaxies in the large  $\Delta$SK bin, \DSFHP\, is positive for all radii ($\sim$ 0.1 dex), even more it shows a positive gradient ($\sim$ 0.07 dex/R$_{\rm eff}$). This suggests that for those galaxies with larger SFR, with respect to the SK-law, the \Ssfr is underestimated from the local \Phyd, particularly at they outskirts.        


Finally, we plot in the bottom panel of  Fig.~\ref{fig:Rad} the radial gradient of \DSFHP\, for galaxies classified according to their vertical distance with respect to the black line in right panel of Fig.~\ref{fig:Sample} (low, average, and high $\Delta$F$_{\rm gas}$; in other words the lack or excess of M$_{\rm mol}$ for a given M$_{\ast}$). For the average $\Delta$F$_{\rm gas}$ the radial distribution of \DSFHP\, is flat and close to zero. For galaxies in the low $\Delta$F$_{\rm gas}$ bin, we find a significant negative radial gradient (\mbox{$\sim$ -0.14 dex/R$_{\rm eff}$}), suggesting that in the outskirts of these galaxies \Ssfr is overestimated by \Phyd. We note a dip of \DSFHP\, at central regions for galaxies in the high $\Delta$F$_{\rm gas}$ bin, inducing a positive gradient ($\sim$ 0.07 dex/R$_{\rm eff}$). These radial trends show the impact of global properties in the local scaling relations such as the \mbox{\Ssfr - \Phyd} one. 

\subsection{Testing the impact of systematic in the \mbox{\Ssfr - \Phyd} relation}
\label{sec:caveats}
\begin{figure}
\includegraphics[width=\linewidth]{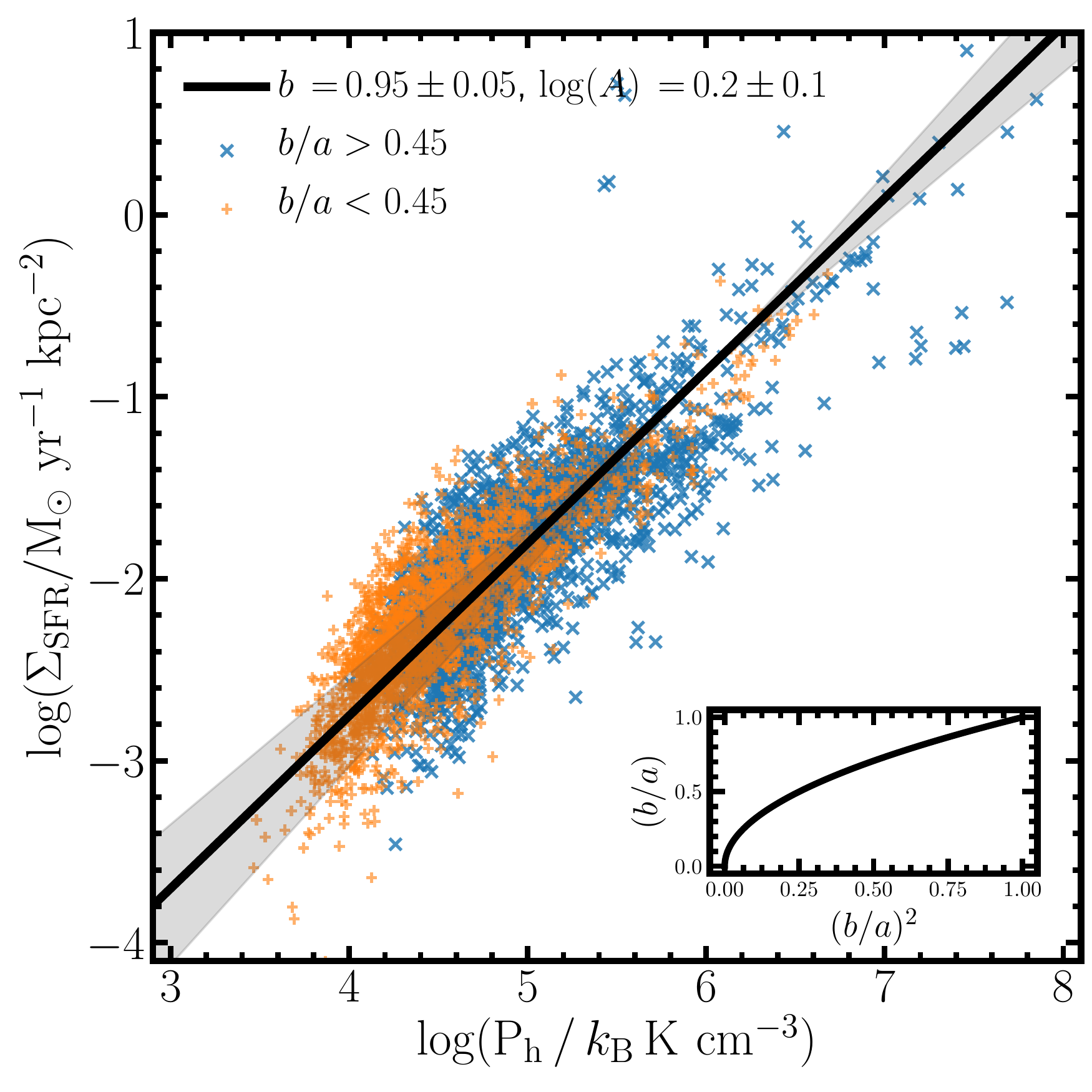}
\caption{The \mbox{\Ssfr - \Phyd} relation,  with regions located in low inclination galaxies ($b/a$ $>$ 0.45) shown with blue crosses and high inclination ones ($b/a$ $<$ 0.45) shown as orange plus symbols. Similar to  Fig.~\ref{fig:SPrel}, the solid line representing and ODR fit to the regions in low-inclined galaxies show a good agreement with the best fit for all the sample in Fig.~\ref{fig:SPrel}. The inset shows the impact of $b/a$ in each axis of the \mbox{\Ssfr - \Phyd} relation.
}
\label{fig:low-inc}    
\end{figure}

In this section we explore the impact of the inclination and a variable \mbox{$\alpha_{\rm CO}$} conversion factor on the \mbox{\Ssfr - \Phyd} relation. Although all the surface densities in this study are corrected by inclination effects, to further explore the impact of galaxy projection on this relation, in Fig.~\ref{fig:low-inc} we plot the \mbox{\Ssfr - \Phyd} relation as in Fig.~\ref{fig:SPrel}, but color-coding galaxies according to inclination: regions located in low inclination galaxies ($b/a$ $>$ 0.45) are indicated with blue crosses and regions located in high inclination galaxies $b/a$ $<$ 0.45)  are indicated with orange plus symbols. We note that on average for low inclined galaxies \Phyd\, and \Ssfr are larger compared to regions located in highly inclined galaxies. We perform a similar analysis as in Sec.~\ref{sec:SPrel} for the low-inclined sample. We find a smaller Pearson correlation coefficient in comparison to the entire sample ($r = 0.80$). 
The ODR fitting shows a similar slope by comparison to the entire sample (\mbox{$b$ = 0.95$\pm$0.05, $A$ = 0.2$\pm$0.1}). The residual of the fit is slightly larger than the one derived from the entire sample ($\sigma$ = 0.30). These results show that inclination does not play a major impact in shaping the slope of the \mbox{\Ssfr - \Phyd} relation. The impact of inclination measured by the $b/a$ ratio for \Ssfr and \Phyd is shown in the inset of Fig.~\ref{fig:low-inc}. This further indicates the impact of inclination in the observed  \mbox{\Ssfr - \Phyd} relation. 

\begin{figure}
\includegraphics[width=\linewidth]{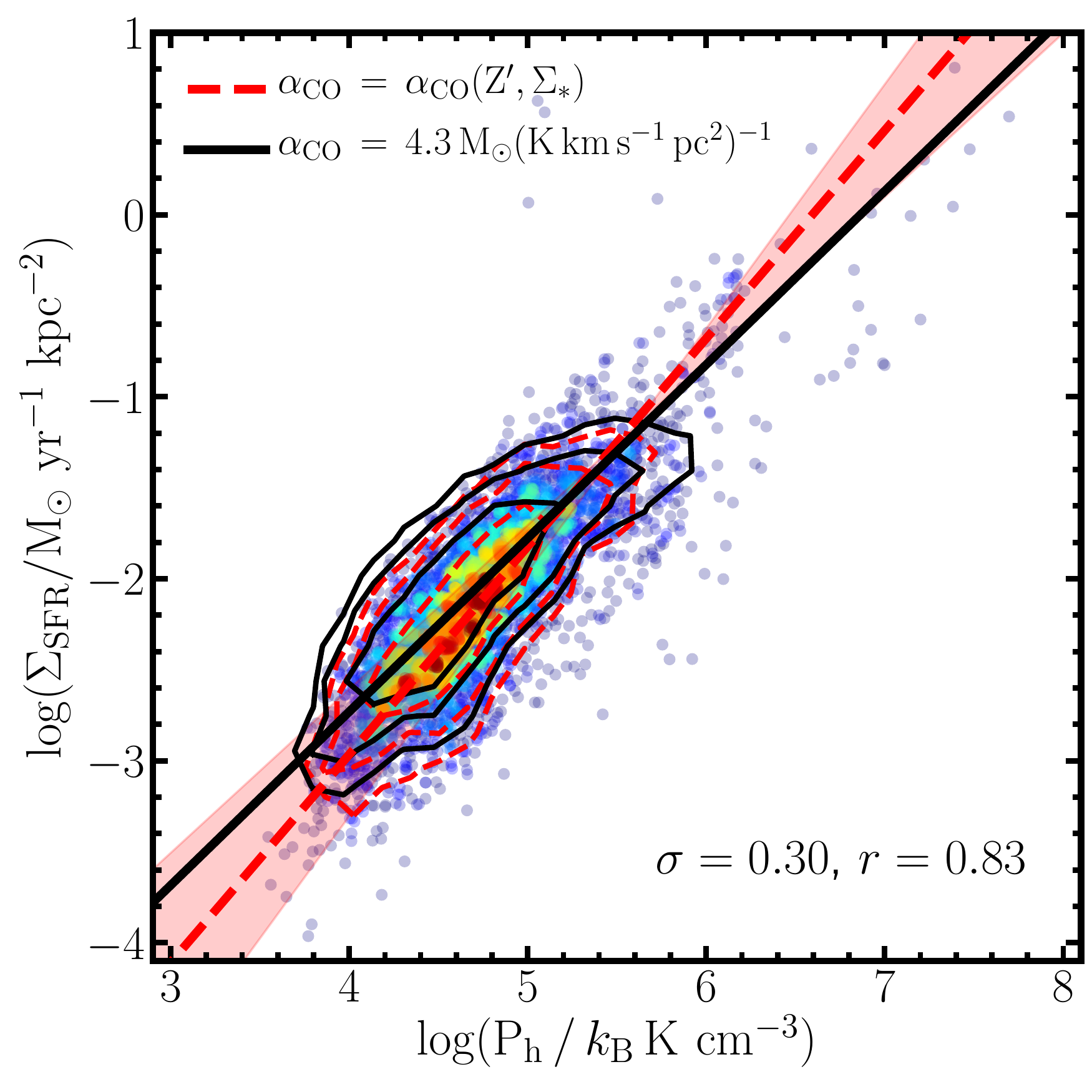}
\caption{The red-dashed contours enclose the \mbox{\Ssfr - \Phyd} relation using a variable  $\mathrm{\alpha_{CO}}$ conversion factor. The best ODR fit is represented by a red-dashed line while the the red shaded area shows the uncertainty of assuming different \Shi densities (see details in Sec.~\ref{sec:Data}). The black contours and black lines are the same as those in Fig.~\ref{fig:SPrel}. By allowing a variable $\mathrm{\alpha_{CO}}$ conversion factor, the slope of the \mbox{\Ssfr - \Phyd} relation is steeper than when adopting a constant factor. 
}
\label{fig:aCO}    
\end{figure}

Along this article, we adopt a constant Milky-Way value of the conversion factor between the CO luminosity and the molecular gas mass density ($\mathrm{\alpha_{CO}}$). Since $\mathrm{\alpha_{CO}}$ can vary, in particular decreasing in regions of high density and temperature where the CO excitation is higher, which tend to be associated with high-pressure inner-galaxy regions \citep[e.g.,][]{Bolatto_2013, Gong_2020}, here we explore how the \mbox{\Ssfr - \Phyd} relation varies when including a variable $\mathrm{\alpha_{CO}}$ (\mbox{$\mathrm{\alpha_{CO}}(Z,\Sstar)$}, see Eq.~\ref{eq:aCO}). In Fig.~\ref{fig:aCO}, datapoints and red-dashed contours represent the \mbox{\Ssfr - \Phyd} relation using \mbox{$\mathrm{\alpha_{CO}}(Z,\Sstar)$} whereas the black contours are the same as those presented in Fig.~\ref{fig:SPrel}. We follow a similar procedure as in Sec.~\ref{sec:SPrel} to derive the best ODR fit to the data. We find that this fit is steeper (red-dashed line, \mbox{$b$ = 1.14$\pm$0.07, $A$ = 0.0$\pm$0.1}) in comparison to the one derived adopting a constant $\mathrm{\alpha_{CO}}$ (black solid line). The slope of the best fit agrees with derivation from numerical simulations of feedback from SN explosions and photoelectric heating \citep[][]{Kim_2013}. However, the scatter of the relation is larger (and the Pearson correlation is smaller, $r =$ 0.83) in comparison to the relation presented in Sec.~\ref{sec:SPrel}.

\subsection{Comparison with the literature}
\label{sec:comp}
\begin{figure}
\includegraphics[width=\linewidth]{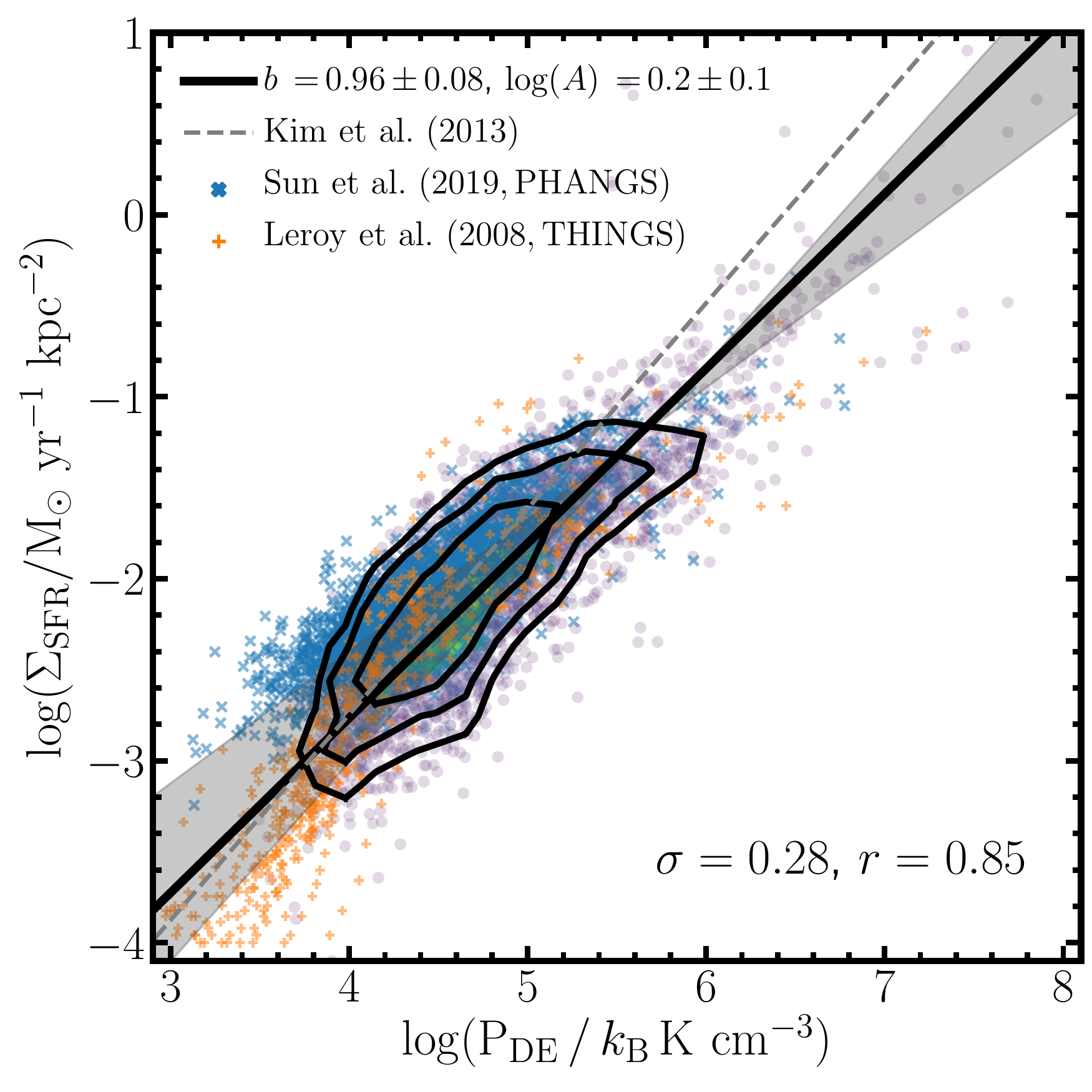}
\caption{
The \Ssfr - \Pde\, relation derived for the EDGE-CALIFA galaxies (contours enclosing 90\%, 80\%, and 50\% of the sample) compared to spatially resolved data from previous studies. Orange plus symbols represent the data from the THINGS survey \citep{Leroy_2008}. Blue crosses represent the data from the PHANGS survey \citep{Sun_2020}. The solid black line represents the best fit to our data. The dashed line shows the prediction from hydrodynamic simulations \citep{Kim_2013}. Despite the differences in observables and samples size, the  \Ssfr - \Pde\, relation for the EDGE-CALIFA sample is in agreement with previous estimates from smaller samples of galaxies.
}
\label{fig:Pde}    
\end{figure}

Using spatially resolved data from surveys including small yet significant samples of galaxies, different studies have shown the tight correlation between \Ssfr and the so-called dynamical equilibrium pressure, \Pde\, (see Eq.~\ref{eq:Pde}). In this section we compare the \Ssfr-\Pde\, relation derived for our dataset of spatially resolved observations of 96 star-forming galaxies from the EDGE-CALIFA survey with those relations derived in the literature. 

In Fig.~\ref{fig:Pde} we plot the \Ssfr-\Pde\, relation for the regions sampled by the EDGE-CALIFA survey. Inwards, the contours enclose 90\%, 80\%, and 50\% of the distribution in this relation. The correlation coefficient of this relation is similar to the one derived for the \mbox{\Ssfr - \Phyd} relation ($r =$ 0.85). Despite the assumptions to derive \Pde, the trend observed of the \mbox{\Ssfr - \Pde} relation with our dataset is in good agreement with those derived in the literature for spatially resolved measurements from the HERACLES+THINGS \citep{Leroy_2008} and PHANGS surveys \citep[][blue x-symbols and orange crosses, respectively]{Sun_2020}. Our sample covers similar values of \Ssfr and \Pde\, than those derived from the PHANGS multi-wavelength dataset, the distribution of our estimate of pressure is slightly shifted towards lower values of \Ssfr\, ($\sim$ 0.2 dex). Although small, this difference could be due, among others to (i) the significant difference between the datasets \citep[][used the photometric data at different wavelengths such as UV, and IR, whereas here we use IFS dataset only in the optical regime]{Sun_2020} in particular this could lead to an underestimation of $\sim$ 0.2 dex when using the dust-corrected \ha luminosity to derive \Ssfr \citep[][although see \citealt{Catalan_Torrecilla_2015}]{Hirashita_2003}, and (ii) we study a large galaxy sample \citep[][used a sample of 28 star-forming galaxies whereas we use a sample of 96 galaxies including massive galaxies with a wider range of morphological types, see Fig.~\ref{fig:Sample}]{Sun_2020}. Furthermore, in Sec.~\ref{sec:PS_ratio} we find an excellent agreement with the best-fit parameters derived by \cite{Sun_2020} when selecting regions with large \ha\, equivalent width ( EW(\ha) $\gtrsim$ 20 \AA).

The HERACLES-THINGS survey, on the other hand, traces the \mbox{\Ssfr - \Pde} relation mainly for galaxies with lower stellar mass than our sample (see Fig.~\ref{fig:Sample}). For a fraction of radial bins (those with similar values of \Pde), the \mbox{\Ssfr - \Pde} relation derived from the this survey is in agreement with those derived in this study. However, there is a significant fraction of radial bins that have smaller  values of \Pde\, in comparison to those estimated from the EDGE-CALIFA or PHANGS surveys. Those radial bins exhibit a sharp drop in their values of \Ssfr as those expected from the previous surveys. It can be the case that for regions with \mbox{$\Pde/ k_{\rm B}\, \lesssim $ 10$^{4}$ $\rm{K\,\,cm^{-3}}$} the \Ssfr does not follows a power-law relation (or with a different index) as the one described for the bulk of our observations.  

Following the same procedure as in Sec.~\ref{sec:SPrel}, we derive the best fit for the \mbox{\Ssfr - \Pde} relation. The best relation from an ODR fit is similar -- within uncertainties -- to the one derived in Sec.~\ref{sec:SPrel} for the \mbox{\Ssfr - \Phyd} relation (\mbox{$b$ = $ 0.96 \pm  0.08$}, \mbox{$\log(A)$ = $ 0.2\pm 0.1$}).  The scatter of this relation is similar to the one derived for the  \mbox{\Ssfr - \Phyd} relation (\mbox{$\sigma$ = 0.28}). The slope of this fit is in agreement with estimations using CO staked spectra from the EDGE-CALIFA survey (Villanueva et al. in prep.). On the other hand, an OLS fitting of this relation is in good agreement with the best-fit derived from the PHANGS dataset (\mbox{$b$ = $ 0.84 \pm  0.04$}, \mbox{$\log(A)$ = $ 0.4\pm 0.1$}). The trend observed for our derivation of the \mbox{\Ssfr - \Phyd} relation agrees with the prediction from a hydrodynamic simulation \citep[$b$ = 1.13, and \mbox{$\log(A)$ = 0.26}, ][]{Kim_2013}. As mentioned by \cite{Sun_2020} using the PHANGS data, our observations also show a shallower slope in comparison to the value expected from the simulation, in particular at large values of \Pde. This difference in slope could reflect systematic effects of the ISM at different locations of the galaxies or the change of the properties for massive galaxies. In Sec.\ref{sec:PS_ratio} we further explore these possibilities. In summary, despite the samples sizes and their differences in their measurement, our estimation of the  \mbox{\Ssfr - \Pde} relation is in good agreement with those presented previously in the literature. 


\subsection{The \Phyd/\Ssfr ratio}
\label{sec:PS_ratio}
%
\begin{figure}
\includegraphics[width=\linewidth]{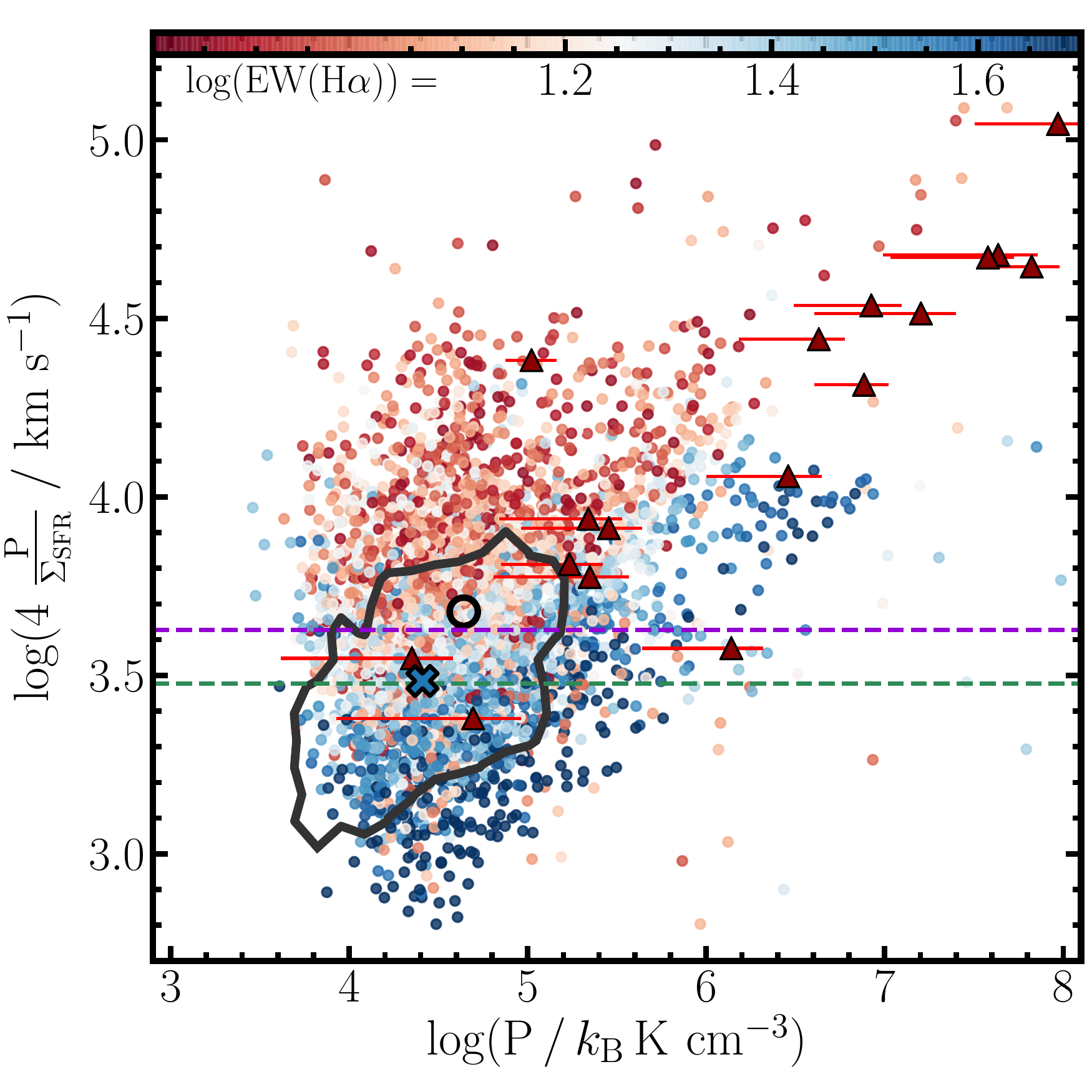}
\caption{The 4 P/\Ssfr ratio against P. The points are color-coded by the EW(\ha) representing the \mbox{4 \Phyd/\Ssfr} vs \Phyd\, relation for the regions sampled in this study. The gray contour encloses 80\% of the 4 \Pde/\Ssfr vs \Pde\, relation derived for regions included in 28 star-forming PHANGS galaxies \citep[][]{Sun_2020}. The empty circle and large blue cross represent the locations of the median ratio and pressure for the previous distributions, respectively. The triangles with error bars represent the integrated measurements from highly turbulent galaxies included in the DYNAMO survey \citep[][]{Fisher_2019}. The dashed green line represents the typical value adopted as the feedback-driven momentum injection per unit mass of stars formed \citep[3000 km s$^{-1}$, ][]{Ostriker_Shetty_2011,Fisher_2019}. The dashed violet line represents the sum of the aforementioned momentum injection with the one derived from the stellar winds and radiation pressure produced by a typical starburst population \citep[$\sim$ 4200 km s$^{-1}$][]{Heckman_2015, Heckman_2017}. Active star-forming regions are in agreement with both the feedback-driven momentum injection and the median values from the literature.  
}
\label{fig:P_SFRratio}    
\end{figure}

From star-formation theory, the ratio between the pressure (P) and the \Ssfr is proportional to the momentum injection per unit mass mostly from supernovae to the ISM, $p_{\ast}/m_{\ast}$  \citep[e.g.,][]{Ostriker_Shetty_2011, Shetty_2012, Faucher-Giguere_2013, Kim_2017}. These studies usually adopt or derive a constant value for this ratio that ranges between \mbox{$\sim 10^3$ - $10^4$ km s$^{-1}$} depending on the adopted conditions for the ISM, the clustering of the supernovae and losses due to interphase mixing \citep[e.g.,][]{Iffrig_2015, Kim_Ostriker_2015, Martizzi_2015, Walch_Naab_2015,Kim_2017, El-Badry_2019, Gentry_2019}. In Fig.~\ref{fig:P_SFRratio} we compare the 4 P/\Ssfr ratio against P, usually this ratio is expressed in units of km s$^{-1}$. The factor 4 comes from assuming that a spherical injection of the momentum flux is centered in the disk mid-plane and that this momentum is converted directly to turbulent pressure \citep{Ostriker_Shetty_2011}; this factor has been directly verified in numerical simulations of disk galaxies \citep[see Fig. 15 of][]{Kim_2013}. In this figure we plot the $4\,\Phyd/\Ssfr$ - \Phyd\, distribution adopting a constant CO conversion factor. Our sample covers a large dynamical range of the $4\,\Phyd/\Ssfr$ ratio (from 5.1 $\times$ 10$^2$ to 1.6 $\times$ 10$^5$ km s$^{-1}$). The median of this ratio is 4.3 $\times$ 10$^4$ km s$^{-1}$ (see empty circle in Fig.~\ref{fig:P_SFRratio}). We also derive this ratio assuming a variable CO conversion factor (see Eq.\ref{eq:aCO}). We find larger values for the $4\,\Phyd/\Ssfr$ ratio when adopting a variable CO conversion factor with a median of $\sim$ 6.4 $\times$ 10$^4$ km s$^{-1}$. On the other hand, we color-coded each region according to their \ha\, equivalent width, EW(\ha). Since this parameter captures the contrast between the adjacent stellar continuum and the \ha\, emission lines, it has been considered as a tracer of the star-formation activity \citep[e.g.,][]{Lacerda_2018, Sanchez_2020ARAA}. For a given \Phyd we find a clear trend, regions with large EW(\ha) have lower values of the $4\,\Phyd/\Ssfr$ ratio. This is particularly evident for regions with the median \Phyd ($\sim $ 10$^{4.4}$ $\rm{K\,\,cm^{-3}}$).

Using the data from \cite{Sun_2020}, we derive the $4\,\Pde/\Ssfr$ - \Pde\, distribution from the PHANGS survey (gray contour enclosing 80\% of their sample). The average $4\,\Pde/\Ssfr$ is smaller than those derived from our regions using \Phyd ($\sim$ 3.0 $\times$ 10$^3$ km s$^{-1}$, see thick large cross in Fig.~\ref{fig:P_SFRratio}). As we mention in Sec.~\ref{sec:comp}, these differences can be expected due to the equations used to estimate \Pde\, and \Phyd, the observables used to derive \Ssfr, and the observed samples of galaxies. Despite these differences, the distribution agrees with our sample, in particular with those regions that show large star-formation activity. In fact the observed trend for the blue regions (i.e., regions with high star-formation activity, EW(\ha) $\gtrsim$ 20 \AA) agrees with the distribution of the $4\,\Pde/\Ssfr$ - \Pde\, ratio from the PHANGS survey. The P/\Ssfr ratio increases with P. A similar trend is also observed for integrated properties of highly-turbulent galaxies included in the DYNAMO survey \citep[][although a shift of $\sim$ -0.2 dex in both axis can be expected due to improved estimations of their \Phyd\, (Girard et al., submitted), the observed trend  holds]{Fisher_2019}.

Following \cite{Kim_Ostriker_2015}, we adopt a fiducial value of \mbox{$\sim$ 3000 km s$^{-1}$} as the value of the $p_{\ast}/m_{\ast}$ expected from supernovae (dashed green horizontal line). Based on the stellar models from \texttt{STARBURST99} \citep[][]{Leitherer_2014} we add to this estimation the momentum injection to the ISM produced by a combination of stellar winds and radiation pressure \citep[$\sim$ 1200 km s$^{-1}$$ $, dashed violet horizontal line,][]{Heckman_2017}. This value should be considered as an upper limit from the stellar models since the momentum flux due to stellar winds for a typical starburst population is smaller and can vary depending on the assumptions of the models \citep[$\sim$ 400 - 700 km s$^{-1}$$ $, e.g.,][]{Ostriker_Shetty_2011,Heckman_2015}. On the one hand, we find that a significant fraction of regions in our sample ($\sim$ 50\%) have a $4\,\Phyd/\Ssfr$ ratio larger than the expected values of momentum injection from models and simulations of star-formation.  On the other hand we find that, when we only consider active star-forming regions (EW(\ha) $\gtrsim$ 20 \AA), the median ratio is consistent with the momentum injection due to SNe explosions ($\sim$ 3.3 $\times$ 10$^3$ km s$^{-1}$). We note that the median $4\,\Pde/\Ssfr$ from PHANGS is in good agreement with the expectation that momentum injection to the ISM is most likely driven by supernovae. 

We recall that one of our selection criteria to ensure that we are only considering star-forming regions in our sample is that their EW(\ha) $>$ 6 \AA\, (see Sec~\ref{sec:Data}). However, we find that for those regions considered as active star-forming (i.e., EW(\ha) $\gtrsim$ 20 \AA), the $4\,\Phyd/\Ssfr$ ratio could indeed be represented by the momentum injection per mass unit expected from SNe, stellar winds and radiation pressure. The trend observed between $4\,\Phyd/\Ssfr$ and \Phyd\, is similar as those derived using (un-)resolved measurements in other star-forming galaxies \citep[][]{Fisher_2019,Sun_2020}. Furthermore, when selecting regions with EW(\ha) $\gtrsim$ 20 \AA, we find a tighter \mbox{\Ssfr - \Phyd} relation ($\sigma$= 0.22 dex) with a strong correlation coefficient ($r$= 0.90) and best-fit ODR parameters very similar as those derived for the PHANGS survey (\mbox{$b$ = $ 0.88 \pm  0.05$}, \mbox{$\log(A)$ = $ 0.5\pm 0.1$}). 

There could be different reasons that explain why we find a significant fraction of star-forming regions with low EW(\ha) with large $4\,\Phyd/\Ssfr$ ratios. We recall that we are convolving the optical maps -- including the \ha map -- to have a common spatial resolution as the CO maps ($\sim$ 7\arcsec). This may induce a dilution of both the \Ssfr and the EW(\ha) for a given value of \Phyd. Thus, for a given sampled area with a low value of EW(\ha) we may be including some active star-forming regions and some others with no significant star-formation (regions with EW(\ha) $<$ 6 \AA, e.g., diffuse ionized gas). On the other hand, in comparison to previous studies exploring the \Ssfr - P relation at (sub-)kpc scales \citep[e.g.][]{Leroy_2008, Sun_2020}, our sample covers a large number of galaxies. This implies that we are probing different regimes of star-formation, including those where star formation is not as intense as those probed previously. This maybe the case for the massive galaxies that we sampled (see Sec.~\ref{sec:global}). It could also be the case that our measurements (i.e., Eq.~\ref{eq:Phyd}) over-estimates the pressure in those regions with EW(\ha) $<$ 20 \AA. This is certainly the case for regions located in structures of galaxies others than the disk like a bulge or a bar. We find that although central regions (i.e., those with  R/R$_{\rm eff} < 0.5$) own the highest pressures ($\sim $ 10$^{5.5}$ $\rm{K\,\,cm^{-3}}$) they exhibit a wide range of $4\,\Phyd/\Ssfr$ ratios and EW(\ha) even more they represent only $\sim$ 17\% of our sample. Thus, those regions with large $4\,\Phyd/\Ssfr$ ratios and low EW(\ha) are not usually located in central regions of galaxies where the bulge dominates. Furthermore, in Sec.~\ref{sec:Disc} we show that barred galaxies have a similar distribution of the $4\,\Phyd/\Ssfr$ ratio in comparison to disk galaxies. From the observational data we suggest that for those regions with small EW(\ha) -- but still considered star-forming regions, 6 $<$ EW(\ha) $<$ 20 \AA -- the feedback provided from \Ssfr may not be sufficient to balance the pressure estimated from the stellar and gas mass densities leading to the observed large $4\,\Phyd/\Ssfr$ ratios. Finally, we note that this analysis assumes that the pressure produced by star formation feedback is mainly due to the specific momentum injected by SNe, $ p_{\ast}/(4m_{\ast})$.  However,  \cite{Ostriker_2010} argued that in regions of low shielding the thermal pressure and magnetic pressure (both driven by feedback) are expected to be comparable to the turbulent pressure, and this has been verified in solar neighborhood simulations by \cite{Kim_Ostriker_2015,Kim_Ostriker_2017}. Other potential sources of pressure associated with star formation, including cosmic rays and radiation, could also contribute to increasing the ratio of pressure to star formation \citep[see e.g.,][]{Ostriker_Shetty_2011, Diesing_2018}. It could be the case that the observed \Phyd/\Ssfr ratios for regions with low EW(\ha) are affected by more sources of feedback other than SNe. 

\section{Discussion}
\label{sec:Disc}
In this study we describe and explore the \mbox{\Ssfr - \Phyd} relation at kpc scales for star forming regions located in a sample of 96 galaxies included in the EDGE-CALIFA survey. In Sec.~\ref{sec:SPrel} we find that this is a tight relation (i.e., with a small scatter, $\sim$ 0.2 dex), with a significant correlation coefficient ($r$ = 0.85, see Fig.~\ref{fig:SPrel}). This highlights the impact of \Phyd\, in shaping \Ssfr at local scales suggesting a scenario in which star-formation activity is self-regulated. As we mention in Sec.~\ref{sec:Intro}, averaged on scales of kpc and Myr, the feedback from SNe and stellar winds from massive stars counteracts the pressure from the gravity produced by the baryonic mass content \citep[e.g.,][]{Thompson_2005, Ostriker_Shetty_2011,Faucher-Giguere_2013}. In this section we discuss the implications of the slope we find for the \mbox{\Ssfr - \Phyd} relation (Sec.~\ref{sec:sub-linear_b}). We also discuss possible explanations for the correlations (or lack thereof) we observe between the residuals of other star-forming scaling relations and those derived from this relation as well as their anti-correlation with stellar properties (Sec.~\ref{sec:resSFR}). Finally, we discuss the impact of global properties on the \mbox{\Ssfr - \Phyd} relation (Sec~\ref{sec:Integ}).

\subsection{The slope of the \mbox{\Ssfr - \Phyd} relation}
\label{sec:sub-linear_b}

Despite the uncertainty in deriving the best \mbox{\Ssfr - \Phyd} relation due to the lack of information regarding the atomic gas distribution for our sampled galaxies, the range of slopes that best describe this relation are slightly sub-linear. In Sec.~\ref{sec:SPrel} we derive the \mbox{\Ssfr - \Phyd} relation assuming an expected range of \Shi densities. The average value of these slopes after 1000 realizations with different \Shi densities is $b$ $\sim$ 0.9. As we find in Sec.~\ref{sec:caveats}, this slope depends on different parameters including the CO conversion factor.

From a theoretical point of view, there are analytical models and numerical simulations of star formation in disk galaxies suggesting that star formation is a self-regulated process. In this scenario, the hydrostatic pressure is balanced by different feedback sources such as turbulent (from stellar winds, supernovae), thermal, magnetic and radiative pressure produced from young stellar objects \citep[e.g.,][]{Ostriker_2010,Ostriker_Shetty_2011, Shetty_2012, Kim_2013, Krumholz_2018}. The \mbox{\Ssfr - \Phyd} relation predicted from these studies is given by $\Sigma_{\rm SFR}  = P/\eta_{\rm tot}$, where $\eta_{\rm tot}$ is the total feedback yield associated with young stars, and the contribution from turbulent pressure, $\eta_{\rm turb} = p_*/(4 m_*)$ is typically the largest single term. (see Sec.\ref{sec:PS_ratio}). Allowing for a dependence of $\eta$ on environment, this yields $\Sigma_{\rm SFR} \propto P^b$, where the power-law index,  $b$ is either unity \citep[for analyical models, e.g., ][]{Ostriker_Shetty_2011} or slightly supra linear \citep[$\sim$ 1.13, for numerical simulations,][]{Kim_2013}. Within the uncertainties, we consider that the \mbox{\Ssfr - \Phyd} relation derived from EDGE-CALIFA galaxies is in agreement with the predicted linear slope from theory of star formation. In Secs.~\ref{sec:SPrel} and \ref{sec:comp} we find that for both estimates of the pressure (\Phyd\, and \Pde) a linear relation describe the bulk of our dataset. However, we observe differences in the slopes depending on the fitting technique we consider. The slopes derived from the OLS fit ($b \sim$  0.84) are flatter than those derived using a ODR fit ($b \sim$ 0.95) for both  the \mbox{\Ssfr - \Phyd} and  \mbox{\Ssfr - \Pde} relations.

As we mention in Sec.~\ref{sec:comp} similar slopes have been observed using different samples and datasets. The slope derived from the spatially resolved dataset from the PHANGS survey using an OLS bisector method the fit is $b \sim$ 0.84 \citep{Sun_2020} while from unresolved measurements from DYNAMO galaxies the slope is also sub-linear \citep[$b \sim $0.76,][]{Fisher_2019}. \cite{Sun_2020} suggested that one possible reason for which they found a sub-linear slope in the \mbox{\Ssfr - \Pde} relation could be the fact that they sampled galaxies/regions for typical star-forming galaxies where the most extreme cases of star formation have not yet been tested. Thus, these author suggested that in order to fully explore the slope of the \mbox{\Ssfr - \Pde} relation,  extreme star-forming regions (or "starburst" regime) such as the central regions of Ultra Luminous Infrared galaxies (ULIRGs) should be also included in the previous analysis. 

From their results using unresolved measurements, \cite{Fisher_2019} suggested different scenarios that could explain the sub-linear slope they found in their sample of highly turbulent galaxies. They consider that the \mbox{\Ssfr - \Phyd} relation can be truly sub-linear, the $p_{\ast}/m_{\ast}$ ratio increases with the pressure, or/and there are other mechanisms that sustain the pressure in disk galaxies. Numerical simulation of star formation regulated by feedback have shown a qualitatively sub-linear relation between the \Ssfr and the pressure \citep{Benincasa_2016}. These simulations suggest that the feedback has a non-linear impact in the scale height of the galaxy. In turn, this affects the second term of Eq.~\ref{eq:Phyd}, specifically the \mbox{$\mathrm{\sigma_{mol}/\sigma_{stars,z}}$} ratio. To estimate \Phyd\, we have adopted -- as most of the observational studies in this regard -- a constant stellar scale height across the galaxy disk which in turn is proportional to their stellar scale length (see Sec.~\ref{sec:Data}). Although the scale height may vary for different position of the disk, we consider that this does not strongly affect the estimation of \Phyd\, and thus the slope \mbox{\Ssfr - \Phyd} relation. In Appendix \ref{app:Phyd} we show that adopting different estimations of the stellar length yield very similar values of the \mbox{$\mathrm{\sigma_{mol}/\sigma_{stars,z}}$} ratio. Furthermore, \cite{Sun_2020} have found that the stellar scale height and length are tightly correlated for their sample of star-forming galaxies, supporting that indeed the scale height can be considered as constant across the galaxy.  

In Sec.~\ref{sec:PS_ratio} we show that the median value of the $4\,\Phyd/\Ssfr$ ratio (an observational proxy for $p_{\ast}/m_{\ast}$) is larger than the fiducial value expected from the momentum flux injection from SNe (see Fig.~\ref{fig:P_SFRratio}). Nevertheless, for most of the regions with high star formation activity (i.e., EW(\ha) $>$ 20\AA), the $4\,\Phyd/\Ssfr$ ratio is in agreement with a momentum injection per unit mass of star formed produced by SNe. For those active star-forming regions we also find that this ratio increases with \Phyd. This trend is also suggested by the $4\,\Pde/\Ssfr$ - \Pde reported by the PHANGS survey \citep[gray contour in Fig.~\ref{fig:P_SFRratio}][]{Sun_2020} as well as the unresolved measurement from DYNAMO galaxies \citep[red triangles with error bars in Fig.~\ref{fig:P_SFRratio}][]{Fisher_2019}. In fact, using those unresolved measurement \cite{Fisher_2019} suggested that this ratio is not constant across the disks of star-forming galaxies increasing with pressure, leading to the sub-linearity found in the \mbox{\Ssfr - \Pde} relation. For our sample we find this is the case, in particular, for regions with large  EW(\ha). This suggests that for those regions the pressure can be inducing a variation of $p_{\ast}/m_{\ast}$ which in turn may play a significant role in shaping the \mbox{\Ssfr - \Pde} relation. Our results from Sec.~\ref{sec:PS_ratio} also suggest that there could be other processes that can induce departures from the $4\,\Phyd/\Ssfr$ ratio expected from the momentum injection from SNe, in particular for regions with low star-formation. Those physical processes which could include magnetic, and/or thermal pressure or cosmic rays may be very relevant to balance the mid plane hydrostatic pressure in those low star-formation regions. 

Different studies have suggested the relevance of other processes than supernovae explosions which can maintain the pressure support in disk galaxies. From their measurements of the relation between the turbulent pressure and \Ssfr, \cite{Sun_2020} suggested that radial inflows induced by structures such as bars or bulges could be another source of turbulent pressure. On the other hand, models that include momentum flux injection to the ISM from other sources such as radiation pressure, photoionization and winds can contribute to the pressure \citep[e.g.,][]{Hopkins_2011,Hopkins_2014,Murray_2011}, although the ray-tracing simulations of \citealt[][see their Fig.12,]{Kim_2018} show that the specific momentum injection from radiation is small compared to that from SNe, and simulations from Lancaster et al (2021, submitted) found that wind momentum contributions are also much smaller than from SNe. In the same direction, recent models of star formation suggest that  radial transport and feedback from supernovae can have similar impact in regulating star formation in disk galaxies \citep{Krumholz_2018}. In any case, these models/simulations are still needed to explain the observed relation between \Ssfr and the pressure at kpc scales for the sample of galaxies in this study. 

Finally, we note that the slope of the \mbox{\Ssfr - \Phyd} relation can vary depending on the assumption to derive the physical quantities. In Sec.~\ref{sec:caveats} we show that the slope of this relation can vary when assuming a variable CO conversion factor ($b \sim$ 1.15). Furthermore, in Sec.~\ref{sec:PS_ratio} we find that the observed correlation between $4\,\Phyd/\Ssfr$ and \Phyd is reduced once we consider a variable CO conversion factor. We suggest that a larger sample of galaxies (covering a wider range of chemical abundances) is require to further understand the impact of the assumption in the derived quantities.

\subsection{The impact of \Phyd\ on other star-forming relations}
\label{sec:resSFR}
\begin{figure}
\centering
\includegraphics[width=\linewidth]{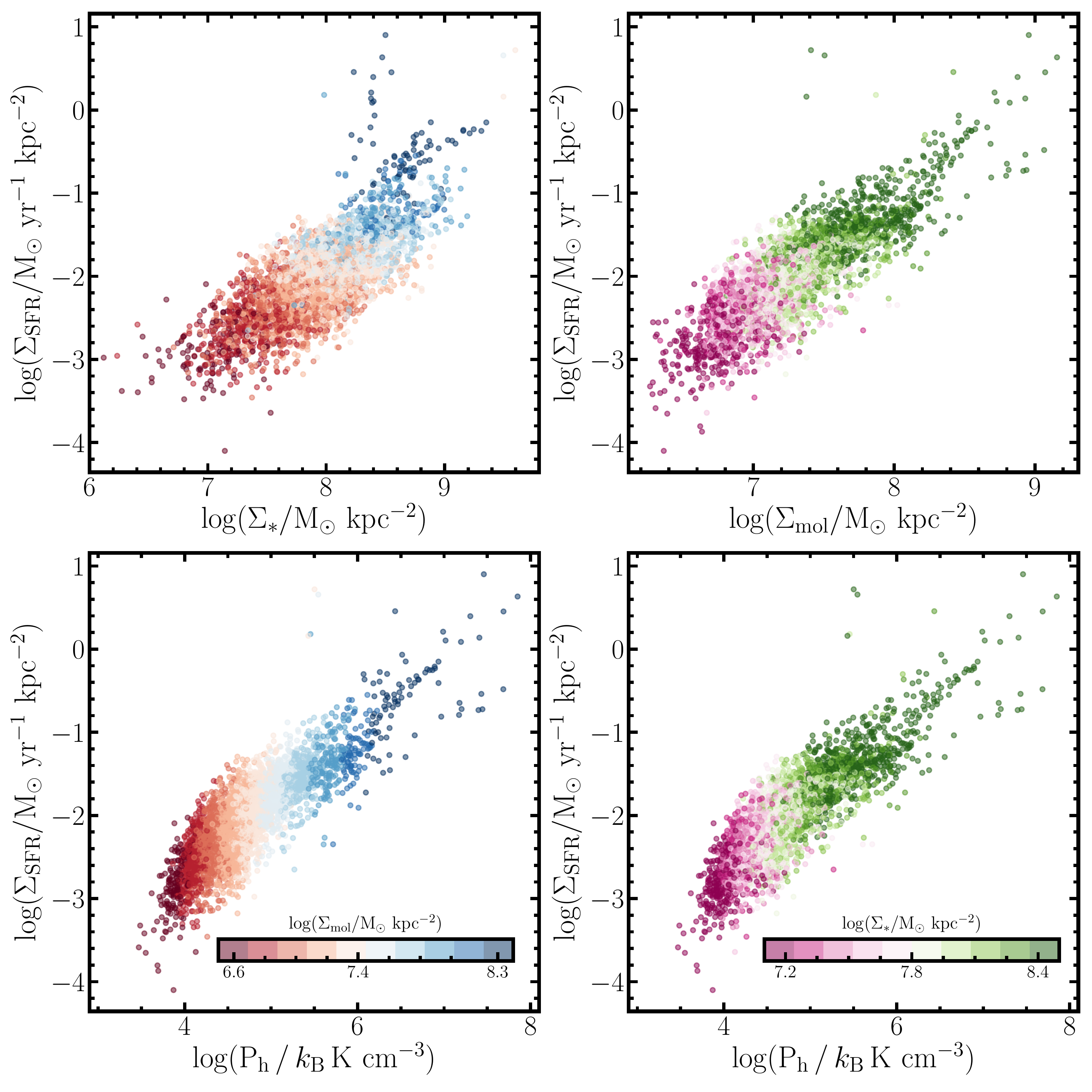}
\caption{Comparison between the resolved star formation main sequence (top-left panel), the resolved Schmidt-Kennicutt law (top-right panel), and the \mbox{\Ssfr - \Phyd} relation (bottom panels). The relations in the left panels are color-coded by the molecular gas mass surface density, \Smol, whereas the right panels are color-coded by the stellar mass surface density. In contrast to the single-variable star-forming scaling relations, the color coding illustrate the minimal impact that either \Sgas or \Sstar has in shaping \Ssfr once \Phyd\, is considered as the independent variable. 
}
\label{fig:SFrelations}    
\end{figure}

In Sec.~\ref{sec:local} we find strong correlations between the residuals of the \mbox{\Ssfr - \Phyd} relation, \DSFHP, and those derived for the resolved star-formation main sequence (\DrSFMS, see Eq.~\ref{eq:rSFMS}) and those from the resolved Schmidt-Kennicutt relation (\DrSFE, see~\ref{eq:rSK}). In contrast, we do not find a strong correlation between the residuals of the resolved molecular gas main sequence (\Drfmol) and \DSFHP (see Fig.~\ref{fig:SPrel_local}). The main goal of this comparison is to quantify whether individual components of the baryonic mass density are driving star formation at kpc scales in the EDGE-CALIFA galaxies.
\cite{Ellison_2020} found a significant correlation between the residuals of the resolved star-forming scaling relation (\DrSFE\, and \DrSFMS) and a secondary correlation with the residuals of the resolved molecular gas main sequence (\Drfmol) and the \DrSFMS. From their results, they suggested that star formation at kpc scales is primarily regulated by the amount of molecular gas, \Smol, with a secondary role for the star formation efficiency, SFE. In the top panels of Fig.~\ref{fig:SFrelations} we illustrate these results. The left panel shows the rSFMS color coded by \Smol. The right panel shows the rSK relation color coded by \Sstar. In both panels the effect that \Smol and \Sstar\, has on each of these relations is clear. For a given value of \Sstar\, (\Sgas), \Ssfr changes with respect to \Sgas (\Sstar). Therefore it is expected to find significant correlation between the residuals of these relations. 

In the bottom panels of Fig.~\ref{fig:SFrelations} we color code the regions in the \mbox{\Ssfr - \Phyd} plane according to their gas and stellar mass densities (left and right panels, respectively). Contrary to the relations in the top panels, these plots show that for a given range of pressures there is no significant change in the values of either \Sgas\, or \Sstar. By construction, \Phyd\, is a combination of both \Sstar\, and \Sgas (see Eq.~\ref{eq:Phyd}). Thus, by including these two terms, the variations in \DSFHP\, are going to scale in a similar way as those observed in the rSFMS or the rSK relation. The vertical distribution of both \Sgas and \Sstar in the \mbox{\Ssfr - \Phyd} plane could also explain the lack of relation between \DSFHP\, and \Drfmol. Previous studies suggested that the \Sgas\Sstar product (or their linear combination in logarithmic scales) better describes \Ssfr than each of them \citep[the so called extended Schmidt-Kennicutt relation, e.g.,][]{Shi_2011,Shi_2018}. Although the expected correlation from these studies has not yet been corroborated using IFS datasets \citep{Lin_2019,Barrera-Ballesteros_2021}, the fact that we find stronger correlation coefficients for the \mbox{\Ssfr - \Phyd} relation (see Sec.\ref{sec:SPrel}), small scatter in comparison to the rSFMS and the rSK (S\'anchez et al., in prep.), and the relations presented in Sec.\ref{sec:local} suggest that the primary driver for the star-formation at kpc scales rather than the individual components of the baryonic mass density is the hydrostatic pressure, \Phyd. 


\subsection{The impact of global properties on the \mbox{\Ssfr-\Phyd} relation}
\label{sec:Integ}

In Sec.~\ref{sec:global} we find that the residuals of the \mbox{\Ssfr - \Phyd} relation apparently correlate with the total stellar mass (see top panel of Fig.~\ref{fig:Mass_Morph}). However, we did not find a strong reduction of the scatter of this relation when we include the total stellar mass as a secondary parameter. We consider that even though for a given \Phyd those regions with high/low \Ssfr tend to be in low-mass/massive galaxies, the pressure is the main parameter that modulates \Ssfr at local scales.  
\begin{figure}
\includegraphics[width=\linewidth]{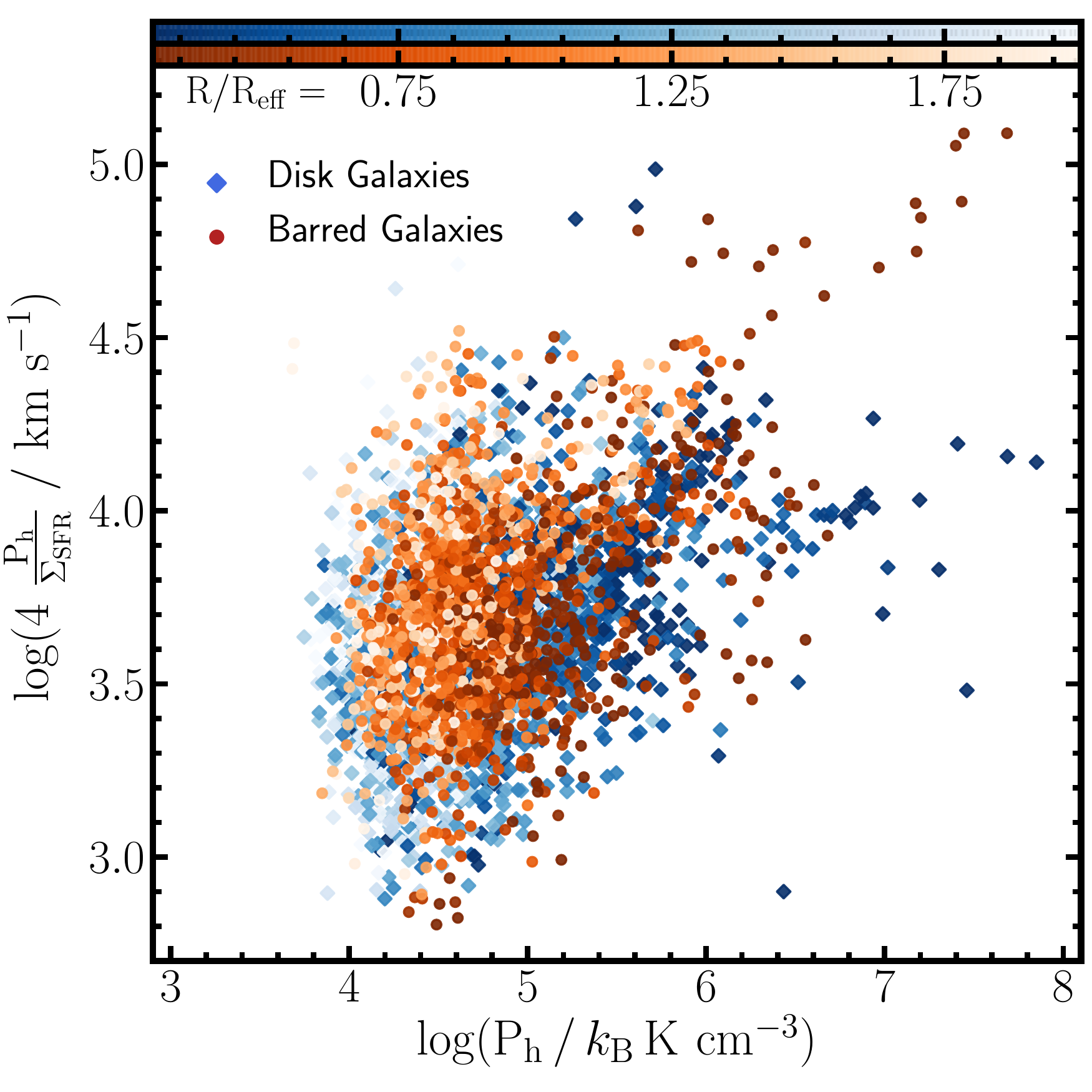}
\caption{Similar to Fig.~\ref{fig:P_SFRratio}. The color of the data points varies according to the galactocentric distance, darker data points indicate regions located closer to center of each galaxy. The circles represent the regions located in barred galaxies whereas the diamonds represent the regions located in disk galaxies. Central regions located in barred galaxies tend to have larger $4\,\Phyd/\Ssfr$ ratios.
}
\label{fig:P_SFRratioBars}    
\end{figure}

We also explore how the \mbox{\Ssfr - \Phyd} relation is affected by the morphology of the host galaxy (see bottom panel of Fig.~\ref{fig:Mass_Morph}). Our analysis shows a mild variation of the residual for different morphological types. We indicate in Sec.~\ref{sec:Data} that most of the targets in our sample are late type galaxies and very few are early-type (see inset in the middle panel of Fig.~\ref{fig:Sample}). For those few early-type galaxies we note that the best \mbox{\Ssfr - \Phyd} slightly overestimated \Ssfr. This may be a hint of the so-called morphological quenching where \Ssfr is halted due to the presence of a bulge rather than the absence of molecular gas \citep[e.g.,][]{Martig_2009,Colombo_2018}. In order to further explore this very interesting possibility, we require a larger sample of galaxies with significant bulge fraction than the one provided in this study. On the other hand, we remind ourselves that the estimation of \Phyd has been derived under the assumption of a thin disk (see Eq.~\ref{eq:Phyd}), therefore it may be not valid for a bulge-dominated galaxies which in turn may lead to this overestimation. 

In Sec.~\ref{sec:sub-linear_b} we suggest that alternative venues that could sustain the pressure other than feedback from supernovae in disk galaxies could be the presence of a bulge or a bar \citep[e.g.][]{Sun_2020}. A significant fraction of our sample includes barred galaxies ($\sim$ 44\%, 45/101) allowing us to test statistically the impact of bars in the estimation of the \mbox{\Ssfr - \Phyd} relation. In Fig.~\ref{fig:P_SFRratioBars}, we compare the regions located in bars and disk galaxies in the \mbox{$4\,\Phyd/\Ssfr$ - \Phyd} plane. Data points are colored by their galactocentric distance with darker points representing regions closer to the center. It is expected that central regions are those with the highest pressures. According to \cite{Sun_2020}, we would expect that those regions with large values of the $4\,\Phyd/\Ssfr$ ratio are located preferentially in barred galaxies. Fig.~\ref{fig:P_SFRratioBars} shows that regions with large values of $4\,\Phyd/\Ssfr$ ratio are located in both barred and disk galaxies. However, there are more regions with \mbox{$4\,\Phyd/\Ssfr$ $\gtrsim$ 10$^{4.5}$ km s$^{-1}$} in barred galaxies than in disk galaxies (20 vs 6). Regions with the highest pressures and the largest \mbox{$4\,\Phyd/\Ssfr$} ratios are located in barred galaxies. Also, the median value of the \mbox{$4\,\Phyd/\Ssfr$} ratio is slightly larger for regions located in barred galaxies than for those located in disk galaxies ($\sim$ 5015 vs. 4500 km s$^{-1}$). Overall, we suggest that bars have a rather mild impact in setting the pressure at kpc scales. Detailed numerical simulations exploring the role of the radial motions as source of pressure are required to quantify these trends. 

 \section{Conclusions}
\label{sec:Conclusions}

The spatially resolved dataset from the EDGE-CALIFA survey  \citep{Bolatto_2017} allow us to estimate the relation between the star formation rate density, \Ssfr and the hydrostatic mid-plane pressure, \Phyd\, for a sample of 4260 star-forming regions located in 96 galaxies of the nearby Universe. This sample covers a significant range of properties which is essential to test the impact of global observables on this spatially resolved scaling relation. The main results of this study are as follows:

\begin{itemize}
    \item We find that \Ssfr strongly correlates with \Phyd\, (Pearson correlation coefficient, $r = 0.84$). This correlation is tight (scatter $\sim$ 0.2 dex). The bulk of this relation is in agreement with a linear relation suggesting that star-formation is an auto-regulated process.
    
    \item From the \mbox{$4\,\Phyd/\Ssfr$} ratios, we suggest that one of the main source of momentum flux injection to the ISM comes from supernovae explosions, in particular for those regions considered as actively star-forming ones (i.e., EW(\ha) $>$ 20 \AA). For those regions with EW(\ha) $<$ 20 \AA, we suggest that either it is required to invoke other sources of pressure such as magnetic, and/or thermal pressure or cosmic rays or that the measured \Ssfr in those regions from the \ha emission line may be polluted with emission not corresponding to pure star-forming regions (e.g., diffuse ionized gas).   

    \item The strong correlation coefficient of the \mbox{\Ssfr - \Phyd} relation in comparison to other star-forming scaling relations (such as the rSFMS and the rSK), the fact that its scatter is very similar to the scatter of those scaling relations, and that its residuals do not correlate with the residuals of the molecular gas main sequence, indicate that \Phyd is probably the main driver of \Ssfr at kpc scales rather than individual components of the baryonic mass. 

    \item Total stellar mass may play a role in shaping the local \Ssfr. For a given \Phyd, \Ssfr decreases as stellar mass increases. However, when imposing a secondary relation with the stellar mass we do not find a strong reduction in the scatter of the \mbox{\Ssfr - \Phyd} relation. If the potential of the host galaxy affects the production of stars at kpc scales, its effect is rather mild.
    
    \item The \mbox{\Ssfr - \Phyd} relation does not seem to be affected by the host galaxy's morphological type. Furthermore, central regions in barred galaxies have similar \mbox{$4\,\Phyd/\Ssfr$} ratios than those located in the center of disk galaxies. In the framework of feedback from recently formed stars, this suggest that bars may play a secondary role as a source of pressure support in late type galaxies.   
\end{itemize}

Our results indicate that thanks to the self-regulation of the star formation, the mid-plane pressure plays a paramount role in shaping the creation of newly born stars at kpc scales in disk galaxies. Injection of momentum flux from supernovae explosions to the ISM is apparently one of the main process that induces this self-regulation. However our analysis, in agreement with previous studies, also suggests that there can be other process that can support the pressure in disk galaxies. Numerical simulations exploring these different channels are thus required as well as spatially resolved observations in actively star-forming regions.

\section*{Acknowledgements}
J.B-B and SFS acknowledge support from the grants IA-100420 and IN100519 (DGAPA-PAPIIT ,UNAM) and funding from the CONACYT grants CF19-39578, CB-285080, and FC-2016-01-1916. SV, ADB, RCL, and VVL acknowledge partial support from NSF-AST1615960. T.W., Y.C., and Y.L. acknowledge support from the NSF through grant AST-1616199. DC acknowlegdes support from the \emph{Deut\-sche For\-schungs\-ge\-mein\-schaft, DFG\/} project number SFB956A.

\section*{Data Availability}
As we mention in Sec.\ref{sec:edgepy}, the data used to derive the physical quantities presented in this article are those available in the \texttt{edge\_pydb} database. A detailed description of the database can be found in Wong et al. (in prep.).


\bibliographystyle{mnras}
\bibliography{main} 

\appendix

\section{Estimation of \mbox{$l_s$}}
\label{app:Phyd}

\begin{figure}
\includegraphics[width=\linewidth]{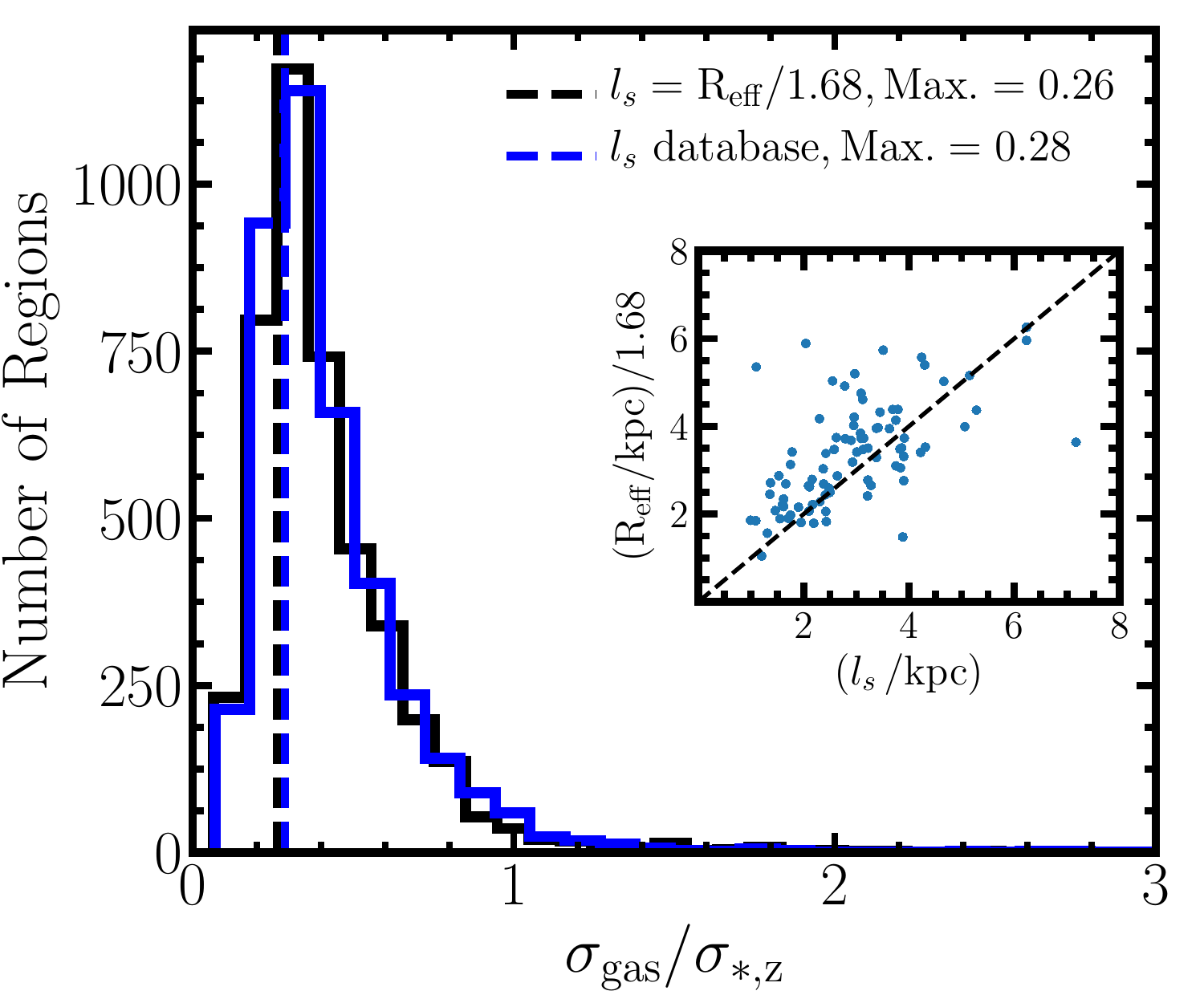}
\caption{Distribution of the \mbox{$\mathrm{\sigma_{mol}/\sigma_{stars,z}}$} velocity dispersion ratio derived for the EDGE-CALIFA survey using the measurements of $l_s$ (blue histogram) and using the adopted value \mbox{$l_s$ = R$_{\rm eff}$/1.68} (black histogram). The blue and black vertical dashed lines represent the peak of each distribution. Both distribution are quite similar, furthermore the inset shows that the two values ($l_s$ and R$_{\rm eff}$/1.68 are quite similar).
}
\label{fig:sigma}
\end{figure}

In Sec.~\ref{sec:Quants} we derive $\sigma_{\ast,z}$ by adopting a relation between the scale height, $h_s$, and the stellar scale length, $l_s$.  This stellar scale length has been measured for this sample of galaxies \citep[][Villanueva et al., in prep.]{Bolatto_2017}. Instead of these measured values, we estimate the scale length as \mbox{$l_s$ = R$_{\rm eff}$/1.68} \cite[i.e., assuming a Sersic profile with $n$ = 1, ][]{Graham_2005}. As we mention in Sec.~\ref{sec:Quants}, we adopt this value to provide an estimation of $l_s$ in larger samples of galaxies where only \mbox{R$_{\rm eff}$} has been determined.  In Fig.~\ref{fig:sigma} we compare the distribution of the \mbox{$\mathrm{\sigma_{mol}/\sigma_{stars,z}}$} ratio using the measurement of $l_s$ included in the \texttt{edge\_pydb} database (blue histogram) and the distribution using the value of $l_s$ adopted in this study (black histogram). We find little difference between these two distributions. The peak of both distribution is very similar ($\sim$ 0.27). In the inset of Fig.~\ref{fig:sigma}, we compare $l_s$ measured and  R$_{\rm eff}$/1.68. These values tend to follow the unity slope. 

\begin{figure}
\includegraphics[width=\linewidth]{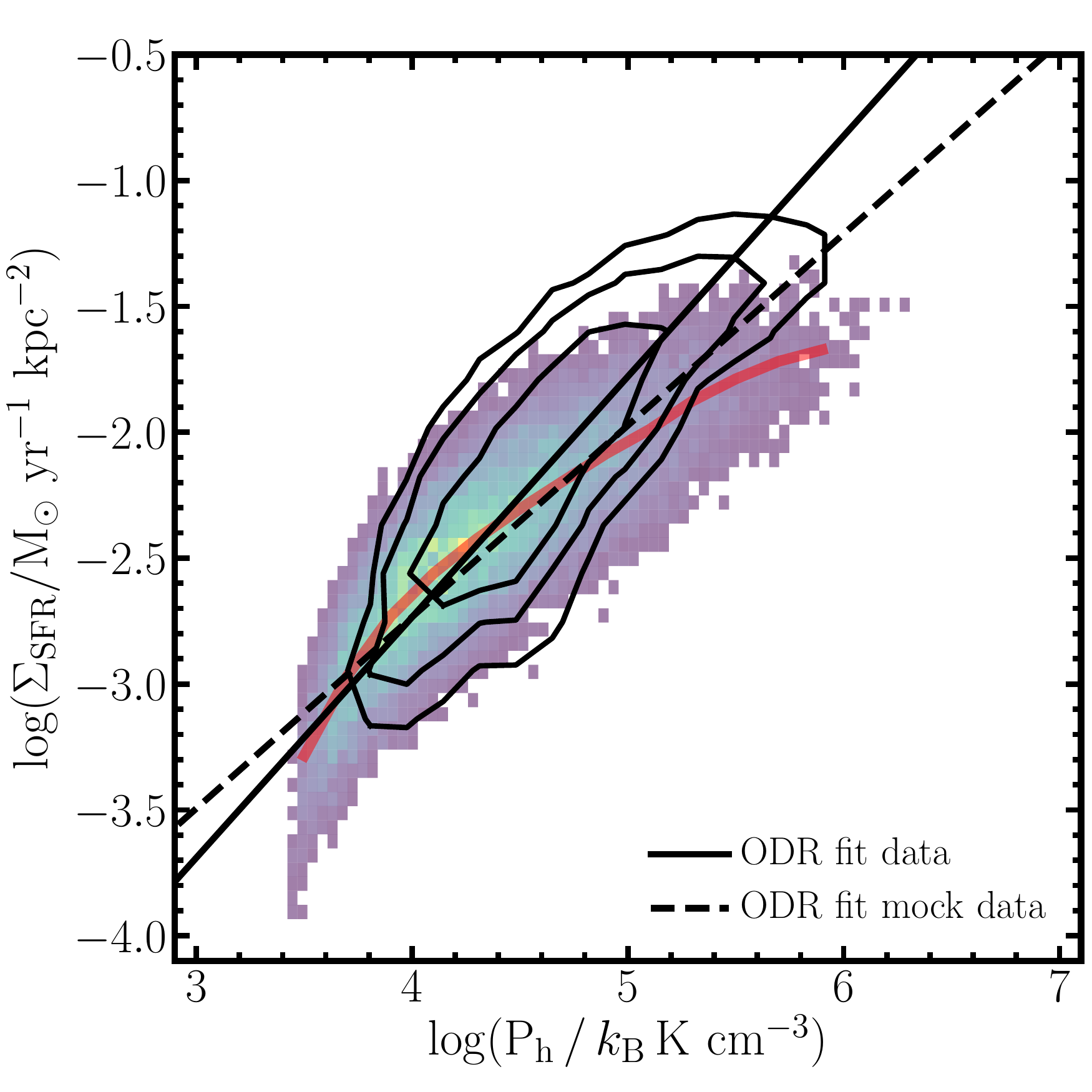}
\caption{The \mbox{\Ssfr - \Phyd} relation derived from the scaling relations presented in S\'anchez et al. (in prep.) for the EDGE-CALIFA sample color coded by the density of points (see details in Appendix \ref{app:Mock}). The red line shows the average \Ssfr for different bins of \Phyd. The contours are the same as those in Fig.~\ref{fig:SPrel}. Black solid line represents the best fit derived from the observed dataset. Black dashed line represents the best relation derived in Sec.~\ref{sec:SPrel}.   
}
\label{fig:P_SFRmock}    
\end{figure}

\section{estimate of the \Ssfr - \Phyd\, relation from scaling relations}
\label{app:Mock}
\begin{figure*}
\includegraphics[width=\linewidth]{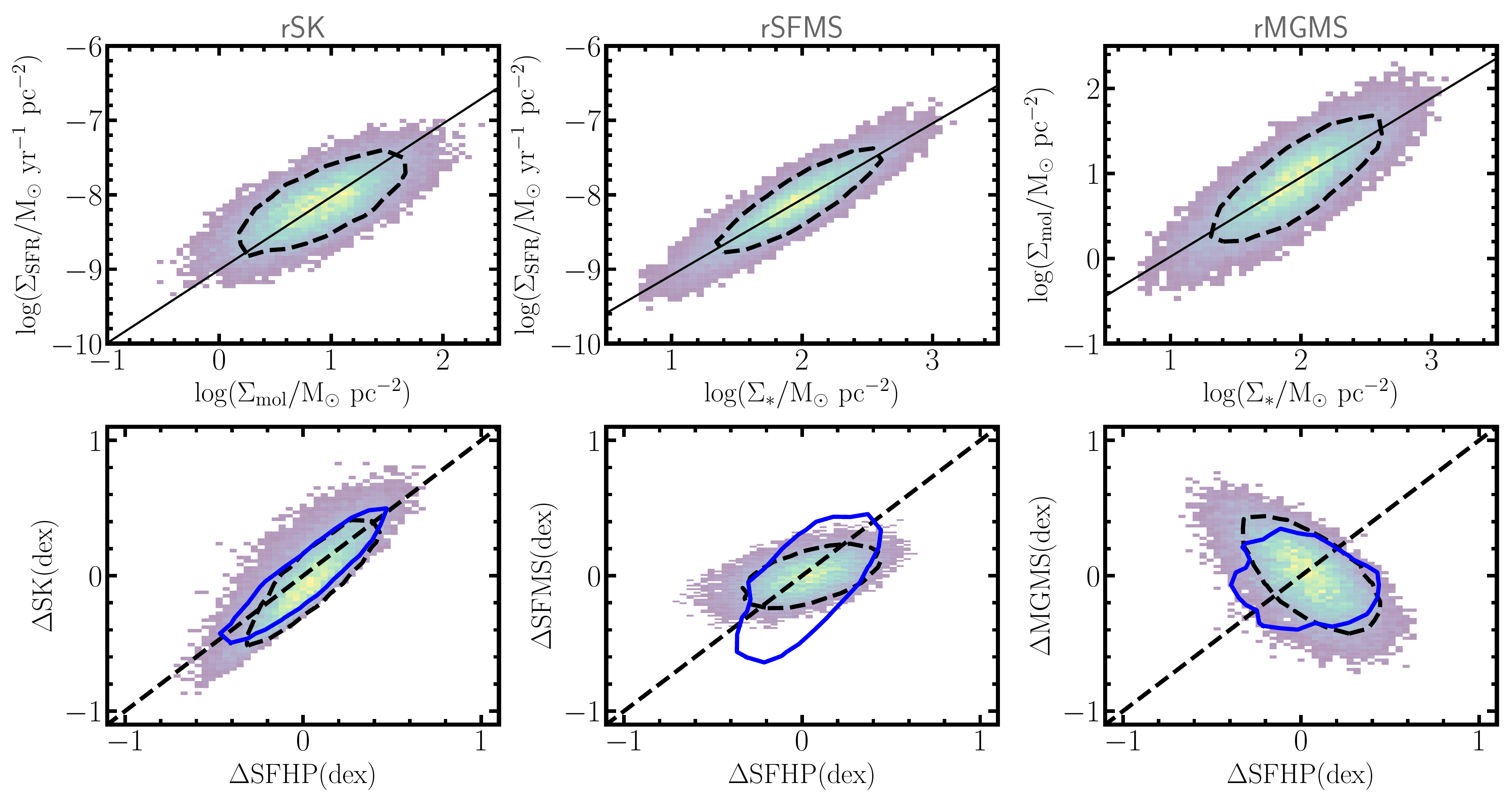}
\caption{({\it top}) Mock distributions of the best scaling relations at kpc scales derived from the EDGE-CALIFA sample. From left to right, the Schmidt-Kennicutt relation (rSK), the star formation main sequence (rSFMS), and the molecular gas main sequence (rMGMS). These relations are derived by assuming the relations presented in Eqs.~\ref{eq:rSFMS}, \ref{eq:rSK}, and \ref{eq:rMGMS}. ({\it bottom}) The residuals of the scaling relations from the top panels against the residuals from the  \Ssfr - \Phyd\, relation derived from these scaling relations. In all panels the dashed black contours include $\sim$ 68\% of the sample, whereas the blue contours in bottom panels represent $\sim$ 68\% of the sample presented in Fig.~\ref{fig:SPrel_local}. The similarity between these distributions allows us to estimate whether they may be induced relations. 
}
\label{fig:RelMock}
\end{figure*}

As we mention in Sec.~\ref{sec:local}, \Ssfr, \Sstar, and \Sgas are closely correlate with each other \citep{Lin_2019}. S\'anchez et al. (in prep.) derive similar scaling relation using the EDGE-CALIFA dataset. Using those scaling relations we derive in this section the \Ssfr - \Phyd\, relation. In Sec.~\ref{sec:local} we present those scaling relations (Eqs.~\ref{eq:rSK}, \ref{eq:rSFMS}, and \ref{eq:rMGMS}). To derive our set of measurements, first we define \Sstar\, as a set of 2$\times$10$^{4}$ values that follow the distribution of the observed stellar density (see S\'anchez et al. for details). This distribution peaks at $\sim$ 10$^{2}$ \msunperpcsq with a dispersion of $\sim$ 0.4 dex. Then, using these values of \Sstar we derive \Ssfr from Eq.~\ref{eq:rSFMS} and \Smol from Eq.~\ref{eq:rMGMS}. We also derive \Ssfr from Eq.~\ref{eq:rSK} using these values of \Smol. We perturb these measurements by adding random noise within a scatter in agreement with the typical uncertainties of each measurement  (i.e., 0.15, 0.28, and 0.20 dex for \Sstar, \Smol, and \Ssfr, respectively). The fiducial values of \Ssfr used in this test are the average of those obtained in Eq.~\ref{eq:rSFMS} and Eq.~\ref{eq:rSK}. In the top panels of Fig.~\ref{fig:RelMock} we show the scaling relations derived in this test. 

Using these values of \Sstar and \Smol we derive \Phyd\, from Eq.\ref{eq:Phyd}. Following the results from Appendix \ref{app:Phyd}, we use the distribution in Fig.~\ \ref{fig:sigma} to random assign the \mbox{$\mathrm{\sigma_{mol}/\sigma_{stars,z}}$} ratio. In Fig.~\ref{fig:P_SFRmock} we show the \Ssfr - \Phyd\, relation from this mock dataset. We follow the same procedure as in Sec.~\ref{sec:SPrel} to derive the best fit to this relation. To compare with the measured \mbox{\Ssfr - \Phyd} relation we overplot the contours from Fig.~\ref{fig:SPrel} as well as the best relation derived in Sec.\ref{sec:SPrel}. The bulk of the distribution from the mocked dataset agrees with observations. However, the trend of the \mbox{\Ssfr - \Phyd} relation from this mock dataset is different from the one derived from observations. On the one hand, for the low-pressure regime, this mock data shows the drop in \Ssfr, consistent with previous measurements from \hi-rich regions \citep[e.g.,][]{Leroy_2008}. On the other hand, in the high pressure end the relation significantly deviates from either the best fit derived from the observations. Even more, the best fit from this mock data is sub-linear (b$\sim$ 0.76 and $\log(A) \sim 0.26$). This test shows that from the local scaling relation there are significant deviations from a linear trend, in particular at the extreme pressures. It also suggests that more mechanisms other than feedback have to be considered in order to properly explain the \mbox{\Ssfr - \Phyd}. We caution that this is not a definitive test. For instance, we are not sampling the scaling relation for dense/starburst regions where it is very likely that they differ from those derived from typical star-forming regions.

In bottom panels of Fig.~\ref{fig:RelMock} we plot the residuals of the scaling relation derived from Eqs.~\ref{eq:rSFMS}, \ref{eq:rSK}, and \ref{eq:rMGMS} against the residuals of the \Ssfr - \Phyd\, relation. For each panel, the cyan and black contours encloses the sample within approximately 1-$\sigma$ of the mock and observed distributions (see, Fig.~\ref{fig:SPrel_local}), respectively. For the star-forming scaling relations (i.e. the rSK and the rSFMS), we find Pearson correlation coefficients among these residuals similar but smaller as those presented in the EDGE-CALIFA dataset ($r = 0.86$, and 0.55, respectively). Contrary to the lack of correlation that we derive in Sec.~\ref{sec:local}, we find in this simulation a significant anti-correlation between the residuals of the rMGMS and the residuals of the \Ssfr - \Phyd\, relation (r = -0.47). 

When we compare the cyan and black contours in the bottom panels of Fig.~\ref{fig:RelMock} we note that for the left panel they are very similar. This can be an indication that the correlation between the residuals of the rSK and the \Ssfr - \Phyd\, relation could be induced by  statistical errors between these scaling relations rather than an physical driven. For the middle panel, we find that distribution of the residuals of the rSFMS and those from the \Ssfr - \Phyd\, relation derived from this simulation are more concentrated around zero than those derived from  observations. This in turn, may suggest that the observed relation of the scatter may have a physical explanation. Finally, from the simulated dataset we are not able to reproduce the lack of correlation between the residuals of the rMGMS and the \Ssfr - \Phyd\, relation. This supports the hypothesis that \Phyd is an observable that describes better the \Ssfr at kpc scales rather than the gas fraction measured by the residuals of the rMGMS. In a more detailed work we explore the implications of studying the residuals of the scaling relations at kpc scales (Sanchez et al., in prep.).  


\bsp	
\label{lastpage}
\end{document}